\documentclass[dvipdfmx,12pt]{article} 
\usepackage{graphicx,amssymb,bbold,bm}
\usepackage{amsmath,amsfonts,epsfig}
\usepackage{braket}
\usepackage[usenames]{color} 
\usepackage{wrapfig}
\usepackage{amsmath,amsfonts,amssymb,bm}
\usepackage{amsfonts,bm}
\usepackage{tikz}
\usepackage{mathtools}

\usepackage{cancel}

\usepackage{ulem}

\newcommand{\p}{\partial}
\newcommand{\pslash}{p\kern-1ex /}
\newcommand{\qslash}{q\kern-1ex /}
\newcommand{\lslash}{l\kern-1ex /}
\newcommand{\sslash}{s\kern-1ex /}
\newcommand{\kaslash}{k_a\kern-2ex /}
\newcommand{\kbslash}{k_b\kern-2ex /}
\newcommand{\Dslash}{{\cal D}\kern-1.5ex /}

\newcommand{\bc}{\overline{c}}

\newcommand{\beqa}{\begin{eqnarray}}
\newcommand{\eeqa}{\end{eqnarray}}

\newcommand{\bpm}{\begin{pmatrix}}
\newcommand{\epm}{\end{pmatrix}}
\newcommand{\bbm}{\begin{bmatrix}}
\newcommand{\ebm}{\end{bmatrix}}

\def\p{\partial}

\newcommand{\Exp}[1]{\left\langle~#1~\right\rangle}
\makeatletter

\@addtoreset{equation}{section}
\makeatother
\usepackage{subfig}
\begin{document}





\def\Tr{\hbox{Tr}}
\newcommand{\be}{\begin{align}}
\newcommand{\ee}{\end{align}}
\newcommand{\bea}{\begin{eqnarray}}
\newcommand{\eea}{\end{eqnarray}}
\newcommand{\beas}{\begin{eqnarray*}}
\newcommand{\eeas}{\end{eqnarray*}}
\newcommand{\nn}{\nonumber}
\newcommand{\bdyg}{\mathcal{G}}
\newcommand{\bdyR}{\mathcal{R}}
\newcommand{\tbdy}{\text{bdy}}
\newcommand{\bmcalT}{\bm{\calT}}
\newcommand{\pExp}[1]{\langle~#1~\rangle}
\newcommand{\calT}{\mathcal{T}}
\newcommand{\bdyK}{\mathcal{K}}
\newcommand{\bdyD}{\mathcal{D}}
\newcommand{\calE}{\mathcal{E}}
\newcommand{\BGbdyD}{\overline{\bdyD}}
\newcommand{\calH}{\mathcal{H}}
\newcommand{\pbdyg}{H}

\def\bena{\begin{eqnarray}}
\def\eena{\end{eqnarray}}

%
\font\cmsss=cmss8
\def\C{{\hbox{\cmsss C}}}
\font\cmss=cmss10
\def\bigC{{\hbox{\cmss C}}}
\def\scriptlap{{\kern1pt\vbox{\hrule height 0.8pt\hbox{\vrule width 0.8pt
  \hskip2pt\vbox{\vskip 4pt}\hskip 2pt\vrule width 0.4pt}\hrule height 0.4pt}
  \kern1pt}}
\def\ba{{\bar{a}}}
\def\bb{{\bar{b}}}
\def\bc{{\bar{c}}}
\def\bphi{{\Phi}}
\def\Bigggl{\mathopen\Biggg}
\def\Bigggr{\mathclose\Biggg}
\def\Biggg#1{{\hbox{$\left#1\vbox to 25pt{}\right.\n@space$}}}
\def\n@space{\nulldelimiterspace=0pt \m@th}
\def\m@th{\mathsurround = 0pt}

\begin{titlepage}

\begin{center}

\vspace{5mm}
{\Large \bf {The higher dimensional instabilities of AdS in holographic semiclassical gravity}} \\ [2pt]

\vspace{5mm}

\renewcommand\thefootnote{\mbox{$\fnsymbol{footnote}$}}
Akihiro Ishibashi${}^{1}$, 
Kengo Maeda${}^{2}$ and 
Takashi Okamura${}^{3}$

\vspace{2mm}

${}^{1}${\small \sl Department of Physics and Research Institute for Science and Technology, } \\ 
{\small \sl Kindai University, Higashi-Osaka, Osaka 577-8502, JAPAN}

${}^{2}${\small \sl Faculty of Engineering, Shibaura Institute of Technology,} \\   
{\small \sl Saitama 330-8570, JAPAN} 

${}^{3}${\small \sl Department of Physics and Astronomy, Kwansei Gakuin University,} \\   
{\small \sl Sanda, Hyogo, 669-1330, JAPAN} 


{\small \tt 
{akihiro at phys.kindai.ac.jp},  {maeda302 at sic.shibaura-it.ac.jp} \\
{tokamura at kwansei.ac.jp}
}

\end{center}


\noindent

\abstract{ 
In the framework of AdS/CFT duality, we consider the semiclassical problem in general quadratic theory of gravity. 
We construct asymptotically global AdS and hyperbolic~(topological) AdS black hole solutions with non-trivial quantum hair in $4$ and $5$-dimensions 
by perturbing the maximally symmetric AdS solutions to the holographic semiclassical equations. 
We find that under certain conditions, our semiclassical solution of hyperbolic AdS black holes can be dynamically unstable against linear perturbations.    
In this holographic semiclassical context, we also study the thermodynamic instability of the hairy solutions in the $5$-dimensional Gauss-Bonnet theory by computing the free energy and show that depending on the parameter of the Gauss-Bonnet theory, the free energy can be smaller than that of the background maximally symmetric AdS solution in both the global AdS and hyperbolic AdS black hole cases.  
} 
\end{titlepage}

\renewcommand\thefootnote{\mbox{\arabic{footnote}}}

\section{Introduction}\label{sec:1}
The semiclassical approximation to quantum gravity is a tractable approach to incorporating quantum effects into gravity.     
In this approach, gravity is treated classically while matter field quantum mechanically through the semiclassical Einstein equations with the source terms given by the vacuum expectation value of the stress-energy tensor for quantum fields. It was shown that within the semiclassical approximation, the Minkowski spacetime is unstable against a certain type of quantum fluctuation~\cite{Horowitz:1978fq, Horowitz:1980fj, Suen:1989bg}. 
One may ask whether such instability is a generic feature of the solutions to the semiclassical Einstein equations. To address this problem, it is desirable to investigate more general curved spacetime cases. However, it is technically difficult to carry out the calculation of the vacuum expectation value of the stress-energy tensor for a quantum field in curved spacetime, while keeping hold the semiclassical Einstein equations (for some exceptional cases, see e.g.~\cite{Starobinsky:1980te, Vilenkin:1985md}). 
 
The AdS/CFT duality~\cite{Maldacena:1997re} provides an efficient method to compute the vacuum expectation value of a strongly coupled quantum field in curved spacetime by using the corresponding dual theory of classical gravity. As first proposed in \cite{Compere:2008us}, 
the semiclassical equations on a $d$-dimensional boundary spacetime can be converted into a boundary condition on a $(d+1)$-dimensional asymptotically AdS bulk spacetime by promoting the boundary metric to a dynamical field. 
In other words, a mixed boundary condition on the AdS boundary represents the semiclassical equations 
on the boundary spacetime.
There have been several works along this line~\cite{Ecker:2021cvz,Natsuume:2022kic,Ahn:2022azl}. 
Recently, the holographic semiclassical problem was explicitly formulated and analytically solved in an asymptotically AdS$_3$ spacetime~\cite{Ishibashi:2023luz}.
It was shown that the maximally symmetric AdS$_3$ spacetime~(or the covering space of the BTZ black hole) is unstable against semiclassical perturbations and also that some new asymptotically AdS$_3$ solutions with quantum hair are thermodynamically unstable~\cite{Ishibashi:2023psu}. 

In this paper, we extend the work~\cite{Ishibashi:2023psu} to the cases of $4$ and $5$-dimensional asymptotically 
AdS spacetime and investigate instabilities of the maximally symmetric AdS spacetime. 
In general $d$-dimensional curved spacetime with $d>3$, there are higher derivative 
corrections in the effective action. So, we solve semiclassical equations in the framework of general 
quadratic theory of gravity coupled with a strongly interacting quantum field with a gravity dual. We first show 
that the semiclassical equations admit the maximally symmetric AdS spacetime in a certain range of 
a given negative cosmological constant. This is because, as we will see, the vacuum expectation value of the stress-energy tensor 
and the other curvature corrections always appear in the semiclassical equations in the form of an effective cosmological constant.  
 
We next construct semiclassical solutions of asymptotically AdS static spacetime with a non-trivial quantum hair 
by perturbing the maximally symmetric AdS spacetime. Our hairy asymptotic AdS solutions are 
classified into two types: (i) asymptotically global AdS solutions obtained from the perturbation of the 
global AdS spacetime, (ii) asymptotically hyperbolic AdS black hole solutions obtained from the perturbation 
of the hyperbolic (zero mass) AdS black hole solution with a Killing horizon. The existence of the solutions depends on the parameters $\alpha_i$~($i=1,2,3$) characterizing the quadratic theory of gravity. In particular, the Einstein gravity with $\alpha_i=0$ admits some hairy solutions in $d=5$ when the gravitational constant $G_5$ is larger than a critical value, while it does not in $d=4$ for any value of the gravitational constant $G_4$. We also show that the $4$ and $5$-dimensional hyperbolic (zero mass) AdS black hole solutions are always dynamically unstable 
when a hairy asymptotic AdS black hole solution exists. This new instability, which is similar to the lower dimensional case~\cite{Ishibashi:2023psu}, is quite different from the well-known semiclassical linear 
instability~\cite{Horowitz:1978fq, Horowitz:1980fj, Suen:1989bg, Simon:1990jn, Simon:1991bm}, 
originating in the higher order derivative terms.     

Finally, we calculate the free energy of the hairy solutions in the $5$-dimensional Gauss-Bonnet gravity theory, 
which is a particular class of the quadratic gravity theory. In the case of both asymptotically global AdS and hyperbolic 
AdS black hole hairy solutions, we find that there is a parameter region in which the free energy of the hairy solution 
is smaller than that of the background maximally symmetric AdS solution, suggesting that a phase transition occurs 
between the hairy solutions and the background ones. In particular, the free energy of the hyperbolic hairy AdS 
black hole solution is smaller than that of the background solution in the Einstein gravity with $\alpha_i=0$.  
 
This paper is organized as follows. In section \ref{sec:2}, we show that the $4$ and $5$-dimensional maximally symmetric AdS 
spacetimes are the solution of the semiclassical equations. In section~\ref{sec:3}, we clarify the parameter range in which 
the hairy static solutions exist, and analytically derive the hairy solutions. In section~\ref{sec:4}, we first provide a general argument of the stability problem against linear perturbations and then show that the $4$ and $5$-dimensional hyperbolic (zero mass) AdS black hole solutions are dynamically unstable. In section \ref{sec:5}, we calculate the free energy of the hairy AdS solutions. Section \ref{sec:6} is devoted to summary and discussions. The notation 
and conventions essentially follow our previous work~\cite{Ishibashi:2023luz}.

\section{The background solution}\label{sec:2}
In this section, we show that the semiclassical Einstein equations admit $d$~($=4,5$)-dimensional maximally symmetric AdS spacetime with curvature length $\ell$ as the background semiclassical solution. In general, there are higher derivative corrections in the effective action for $d>3$~\cite{BirrellDavies}. In this paper, we consider the quadratic gravity to incorporate the effect. As shown below, the curvature corrections appear as an effective cosmological constant in the semiclassical Einstein equations, and hence the curvature length $\ell$ is different from that of the (bare) negative cosmological constant $\Lambda_d$. 
  
\subsection{The set up}\label{subsec:2}
\begin{figure}[htbp]
  \begin{center}
\includegraphics[width=125mm]{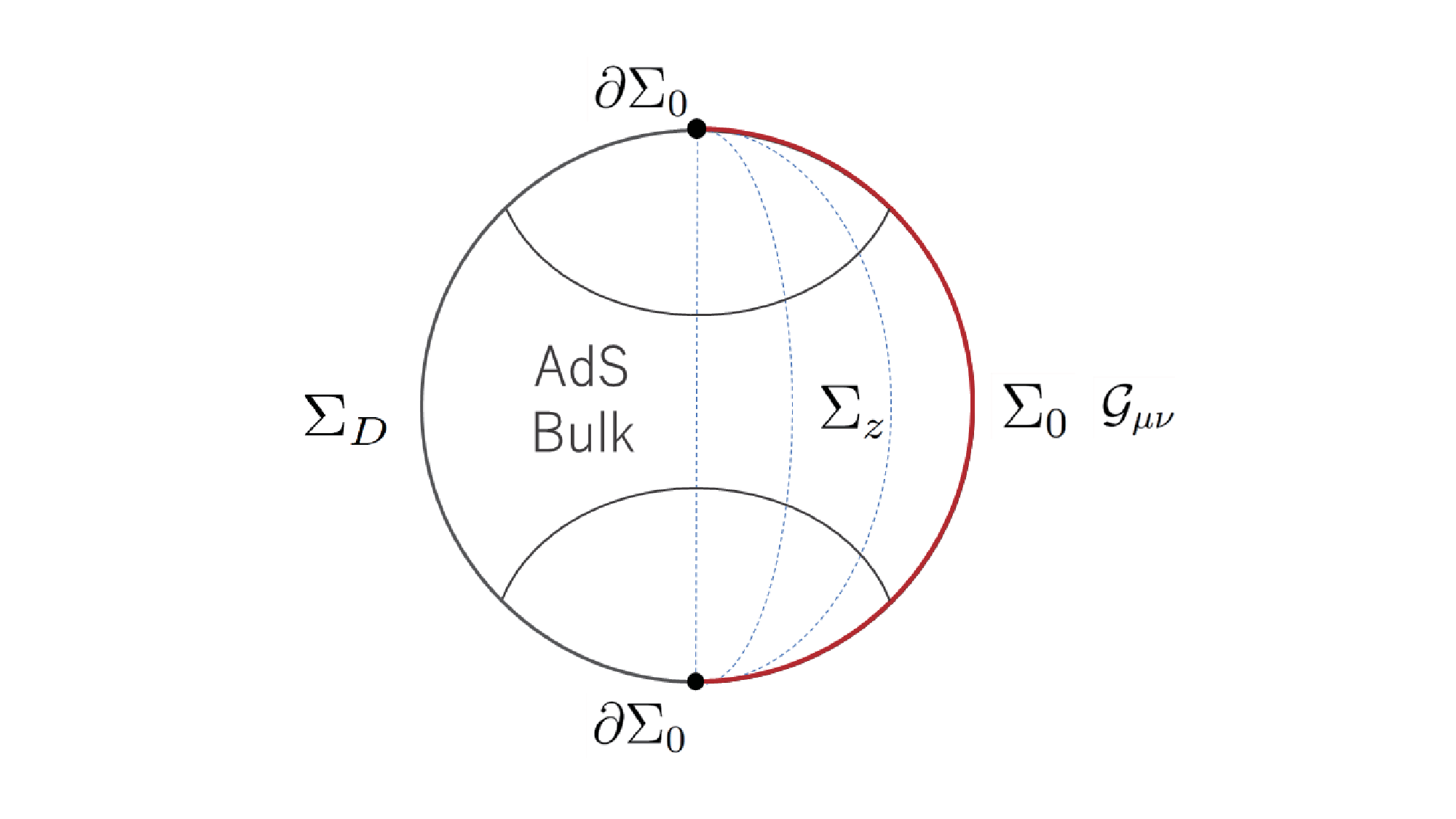}  
  \end{center}
  \caption{{\small A time-slice of the (conformally compactified) AdS bulk spacetime foliated by $z=const$. hypersurfaces $\Sigma_z$, each of $\Sigma_z$ (denoted by the {\it dotted curve}) itself is an AdS spacetime one dimensional lower than the bulk AdS. The conformal boundary of the bulk AdS is divided into the left-part $\Sigma_D$ and the right-part $\Sigma_0$, and these two are matched at the corner $\partial \Sigma_0$. On the-right part $\Sigma_0$, the boundary metric is supposed to satisfy the holographic semiclassical Einstein equations, or in other words the mixed boundary condition is imposed on the bulk metric. For definiteness, we assume that on the left-part $\Sigma_D$, the Dirichlet boundary conditions are imposed on the bulk metric. When $\Sigma_0$ includes a $d$-dimensional boundary black hole, the bulk spacetime also includes a 
$d+1$-dimensional black hole with a horizon inside the bulk. The two {\it hyperbolic curves} denote one of the 
possible bulk horizons. }}
  \label{conformal_ST}
 \end{figure}

According to the AdS/CFT duality~\cite{Maldacena:1997re}, the vacuum expectation value of the stress-energy 
tensor $\Exp{\calT_{\mu\nu}}$ is derived from the $d+1$-dimensional bulk gravity action. We start with the bulk metric 
\begin{align}
\label{bulk_metric}
 ds_{d+1}^2&=G_{MN}dX^MdX^N \nonumber \\
&=\Omega^{-2}(z)dz^2+g_{\mu\nu}(z,x)dx^\mu dx^\nu \nonumber \\
&=\Omega^{-2}(z)(dz^2+\tilde{g}_{\mu\nu}(z,x)dx^\mu dx^\nu), \qquad \Omega(z):=\frac{\ell}{L}\sin\frac{z}{\ell}, 
\end{align}
where $X^M=(z,\,x^\mu)$ and $L$~($\ell$) is the bulk~(boundary) AdS length. The conformal 
boundary metric $\bdyg_{\mu\nu}$ is defined by 
\begin{align}
\label{def_bd_metric}
\bdyg_{\mu\nu}(x):=\lim_{z\to 0}\Omega^2(z)G_{\mu\nu}(z,\,x)=\lim_{z\to 0}\tilde{g}(z,\,x). 
\end{align}  
Here, we assume that the AdS$_{d+1}$ bulk spacetime is foliated by a family of the $z=\text{const.}$ 
hypersurface and the limit hypersurface $\Sigma_0:=\lim_{z\to 0}\Sigma_z$ approaches a portion of the 
conformal boundary $\p M$, as shown in Fig.~\ref{conformal_ST}. 
We consider the $d$-dimensional spacetime with the conformal boundary metric $\bdyg_{\mu\nu}$ 
satisfying the semiclassical Einstein equations. Following~\cite{Ishibashi:2023luz}, we shall 
impose the Dirichlet boundary condition at the other part of the conformal boundary 
$\Sigma_D:=\p M\backslash\{\Sigma_0\cup\p\Sigma_0  \}$~(see Fig.~\ref{conformal_ST}).

The $d+1$-dimensional bulk action consists of the 
$d+1$-dimensional Einstein-Hilbert action $S_\text{EH}$, the $d$-dimensional 
Gibbons-Hawking term $S_\text{GH}$ and the counter term $S^{(d)}_\text{ct}$ as  
\begin{align}
\label{bulk_action}
 S_\text{bulk}&=S_\text{EH}+S_\text{GH}+S_\text{ct} \nonumber \\
&=\int \frac{d^{d+1}X\sqrt{-G}}{16\pi G_{d+1}}\left(R(G)+\frac{d(d-1)}{L^2}  \right)
+\int \frac{d^dx\sqrt{-g}}{8\pi G_{d+1}}K \Bigl{|}_{z=0} +S^{(d)}_\text{ct}. 
\end{align}
In $d=4,\,5$ cases, $S^{(d)}_\text{ct}$ is explicitly written by the 
Ricci curvature $R_{\mu\nu}(g)$ of the metric $g_{\mu\nu}$ as
\begin{subequations}
\begin{align}
\label{bulkct_4}
&S^{(4)}_\text{ct}=-\int \frac{d^4x\sqrt{-g}}{16\pi G_5}\left(\frac{6}{L}+\frac{L}{2}R(g)
-\frac{L^3}{8}\left\{R_{\mu\nu}(g)R^{\mu\nu}(g)-\frac{1}{3}R^2(g)\right\} \ln \frac{z^2}{4\ell^2}  \right), 
\\
\label{bulkct_5}
&S^{(5)}_\text{ct}=-\int \frac{d^5x\sqrt{-g}}{16\pi G_6}\left(\frac{8}{L}+\frac{L}{3}R(g)
+\frac{L^3}{9}\left\{R_{\mu\nu}(g)R^{\mu\nu}(g)-\frac{5}{16}R^2(g)\right\}  \right). 
\end{align}
\end{subequations}

The expectation value of the stress-energy tensor $\Exp{\calT_{\mu\nu}}$ is derived from 
the variation of the bulk on-shell action~\cite{deHaro:2000vlm, Balasubramanian:1999re}. As shown in 
Ref.~\cite{Ishibashi:2023luz}, it is written in terms of the 
extrinsic curvature $\tilde{K}_{\mu\nu}$ with respect to the metric $\tilde{g}_{\mu\nu}$ as
\begin{align}
\label{vev_SE_tensor}
 \Exp{\calT_{\mu\nu}}
 &={-}\dfrac{2}{\sqrt{-\bdyg}}\frac{\delta S_\text{bulk}}{\delta \bdyg^{\mu\nu}}
\nonumber \\
&=\lim_{z\to 0}\frac{1}{8\pi G_{d+1}L}\Biggl{[\dfrac{L^2}{(d-2)\Omega^{d-2}}
\Bigl\{\tilde{K}\tilde{K}_{\mu\nu}-\dfrac{\tilde{g}_{\mu\nu}}{2}(\tilde{K}^{\alpha\beta}
\tilde{K}_{\alpha\beta}+\tilde{K}^2})\Bigr\}
\nonumber \\
&\hspace*{0.5truecm}
-L\tilde{g}_{\nu\rho}\left(\frac{L\Omega}{d-2}\frac{\p}{\p z}+1\right)
\frac{{\tilde{K}_\mu}^\rho-{\delta_\mu}^\rho\tilde{K}}{\Omega^{d-1}}
- \frac{(d-1)\tilde{g}_{\mu\nu}}{2\Omega^d}
       (1-L\Omega')^2  
  \Biggr]
  +\tau^{(d)}_{\mu\nu} \,,  
\end{align}
where the {\it prime} \lq\lq\ $'$ \rq\rq\ denotes the $z$-derivative, 
\begin{align}
\label{def_tilextr}
\tilde{K}_{\mu\nu}=-\frac{1}{2}\p_z\tilde{g}_{\mu\nu}
\end{align}
and 
$\tau^{(d)}_{\mu\nu}$ is given by 
\begin{subequations}
\begin{align}
\label{vev_SE_ct_d=4}
\tau^{(4)}_{\mu\nu}&=-\frac{L^3}{64\pi G_5}\,\Biggl\{\left(\tilde{D}^2-\frac{2}{3}\tilde{R}\right)
\tilde{R}_{\mu\nu}-\frac{1}{3}\tilde{D}_\mu \tilde{D}_\nu \tilde{R}+2\tilde{R}_{\alpha\mu\beta\nu}
\tilde{R}^{\alpha\beta} \nonumber \\
&\hspace*{2.0truecm}
-\frac{\tilde{g}_{\mu\nu}}{2}
\left(\tilde{R}_{\alpha\beta}\tilde{R}^{\alpha\beta}+\frac{1}{3}\tilde{D}^2\tilde{R}
-\frac{1}{3}\tilde{R}^2\right)  \Biggr\}\ln\frac{z^2}{4\ell^2}, 
\\
\label{vev_SE_ct_d=5}
\tau^{(5)}_{\mu\nu}&=\frac{L^3}{72\pi G_6\Omega}\Biggl\{\left(\tilde{D}^2-\frac{5}{8}\tilde{R}\right)\tilde{R}_{\mu\nu}
-\frac{3}{8}\tilde{D}_\mu \tilde{D}_\nu \tilde{R}+2\tilde{R}_{\alpha\mu\beta\nu}\tilde{R}^{\alpha\beta}
\nonumber \\
&\hspace*{2.0truecm}
-\frac{\tilde{g}_{\mu\nu}}{2}
\left(\tilde{R}_{\alpha\beta}\tilde{R}^{\alpha\beta}+\frac{1}{4}\tilde{D}^2\tilde{R}
-\frac{5}{16}\tilde{R}^2\right)  \Biggr\}. 
\end{align}
\end{subequations}
Here, $\tilde{R}_{\mu\nu}$ is the Ricci tensor of the conformal metric $\tilde{g}_{\mu\nu}$
and the on-shell condition 
\begin{align}
\label{bulk_Einstein}
R_{MN}-\frac{1}{2}G_{MN}R-\frac{d(d-1)}{2L^2}G_{MN}=0
\end{align}
is imposed to derive (\ref{vev_SE_tensor}).  

In the $d$-dimensional boundary theory, we consider the following gravity action with the 
quadratic curvature corrections, 
\begin{subequations}
\begin{align}
\label{bd_action}
&{\mathcal S}={\mathcal S}_\text{bd}+{\mathcal S}_\text{GH}+{\mathcal S}_\text{ct}
+S_\text{bulk}, 
\\
& {\mathcal S}_\text{bd}=\int \frac{d^dx\sqrt{-\bdyg}}{16\pi G_d}
(\bdyR+\alpha_1\bdyR^2+\alpha_2\bdyR_{\mu\nu}\bdyR^{\mu\nu}
+ \alpha_3\, \bdyR_{\mu\nu}^{\alpha\beta}\, \bdyR^{\mu\nu}_{\alpha\beta}
-2\Lambda_d), 
\end{align}
\end{subequations}
where ${\mathcal S}_\text{GH}$, ${\mathcal S}_\text{ct}$,
$\bdyR_{\mu\nu}^{\rho\sigma} := \bdyR_{\mu\nu}{}^{\rho\sigma}
= \bdyR^{\rho\sigma}{}_{\mu\nu}$,
and $\bdyR_{\mu\nu}$ are the $(d-1)$-dimensional 
(generalized) Gibbons-Hawking surface term, the counter term,
the Riemann tensor,
and the Ricci tensor of the conformal 
metric $\bdyg_{\mu\nu}$, respectively, and where $\Lambda_d$ is the (bare) 
cosmological constant, $\alpha_i~(i=1,2,3)$ are free parameters of the quadratic gravity theory. 

By taking variation with respect to $\bdyg_{\mu\nu}$ of the action~(\ref{bd_action}),  
one obtains the semiclassical Einstein equations 
\begin{subequations}
\begin{align}
  & \calE_{\mu\nu} = 8\, \pi\, G_{d}\, \Exp{\calT_{\mu\nu}}, 
\label{semi_Eqs} \\
  & \calE_{\mu\nu}
  := \bdyR_{\mu\nu}-\frac{\bdyR}{2}\bdyg_{\mu\nu}+\Lambda_{d}\,\bdyg_{\mu\nu}
  + \alpha_1 \mathcal{H}^{(1)}_{\mu\nu}
  +\alpha_2 \mathcal{H}^{(2)}_{\mu\nu}+\alpha_3 \mathcal{H}^{(3)}_{\mu\nu}
  ~,
\label{eq:def-calE}
\end{align}
\end{subequations}
where $\mathcal{H}^{(i)}_{\mu\nu}~(i=1,2,3)$ are given by 
\begin{subequations}
\begin{align}
\label{def:H}
& \mathcal{H}^{(1)}_{\mu\nu}=2(\bdyR_{\mu\nu}-\bdyD_\mu\bdyD_\nu)\bdyR-
\bdyg_{\mu\nu}\left(\frac{1}{2}\bdyR^2-2 \bdyD^2 \bdyR   \right), 
\\
& \mathcal{H}^{(2)}_{\mu\nu} =2\bdyR_{\mu\rho\nu\sigma}\bdyR^{\rho\sigma}
+\bdyD^2 \bdyR_{\mu\nu}
-\bdyD_\mu\bdyD_\nu\bdyR
-\frac{1}{2}\bdyg_{\mu\nu}(\bdyR_{\sigma\rho}\bdyR^{\sigma\rho}- \bdyD^2\bdyR), 
\\
& \mathcal{H}^{(3)}_{\mu\nu}=2\bdyR_{\mu\rho\sigma\tau}{\bdyR_\nu}^{\rho\sigma\tau}
- \frac{\bdyg_{\mu\nu}}{2}\, \bdyR_{\alpha\beta}^{\rho\sigma}\,
  \bdyR^{\alpha\beta}_{\rho\sigma}
+4{\bdyR}_{\mu\rho\nu\sigma}\bdyR^{\rho\sigma}
-4{\bdyR}_{\mu\rho}{\bdyR^{\rho}}_\nu
-2\bdyD_\mu\bdyD_\nu \bdyR+4 \bdyD^2 \bdyR_{\mu\nu}. 
\end{align}
\end{subequations}
\subsection{$d=4$ background solution}
For the background semiclassical solution of Eqs.~(\ref{semi_Eqs}), we consider the maximally 
symmetric spacetime with the conformal boundary metric $\overline{\bdyg}_{\mu\nu}$ satisfying 
\begin{align}
\label{back_Riebdyg}
\overline{\bdyR}_{\mu\nu\alpha\beta}=
-\frac{1}{\ell^2}(\overline{\bdyg}_{\mu\alpha}\overline{\bdyg}_{\nu\beta}
-\overline{\bdyg}_{\mu\beta}\overline{\bdyg}_{\nu\alpha}). 
\end{align}
In $d=4$ case, $\overline{ \Exp{\calT_{\mu\nu}} }$ in Eq.~(\ref{vev_SE_tensor}) takes a nonzero value 
\begin{align}
\label{vev_d=4_bg_SE}
\overline{ \Exp{\calT_{\mu\nu}} }
=-\frac{3L^3\overline{\bdyg}_{\mu\nu}}{64\pi G_5\ell^4}, 
\end{align} 
due to the existence of the Weyl anomaly. Since ${\cal H}^{(i)}_{\mu\nu}~(i=1,2,3)=0$ for 
the metric $\overline{\bdyg}_{\mu\nu}$ satisfying Eq.~(\ref{back_Riebdyg}), the semiclassical Eqs.~(\ref{semi_Eqs}) 
reduce to 
\begin{align}
\label{cosmo_const_d=4}
& \Lambda_4=-\frac{3}{\ell^2}\left(1+\frac{\pi\gamma_4}{8}  \right), 
\end{align}
where $\gamma_d$ is the dimensionless parameter~\cite{Ishibashi:2023luz} 
\begin{align}
\label{def_gamma_para}
\gamma_d:=\frac{G_dL}{\pi G_{d+1}}\left(\frac{L}{\ell}\right)^{d-2}, 
\end{align}
which determines the magnitude of the backreaction of $\Exp{\calT_{\mu\nu}}$ on the geometry 
through the semiclassical Eqs.~(\ref{semi_Eqs}).  
By Eq.~(\ref{cosmo_const_d=4}) the bare cosmological constant $\Lambda_4$ is smaller than the 
effective cosmological constant $-3/\ell^2$, and the solution always exists for an arbitrary negative value 
$\Lambda_4$ by choosing a suitable length $\ell$. 
\subsection{$d=5$ background solution}
In $5$-dimensional case,
$\overline{ \Exp{\calT_{\mu\nu}} }$
in Eq.~(\ref{vev_SE_tensor}) is zero, as the Weyl anomaly vanishes. So, substituting 
\begin{align}
\label{bg_d=5_H}
{\cal H}^{(1)}_{\mu\nu}=-\frac{40}{\ell^4}\overline{\bdyg}_{\mu\nu}, \quad 
{\cal H}^{(2)}_{\mu\nu}=-\frac{8}{\ell^4}\overline{\bdyg}_{\mu\nu}, \quad 
{\cal H}^{(3)}_{\mu\nu}=-\frac{4}{\ell^4}\overline{\bdyg}_{\mu\nu}  
\end{align}
into the semiclassical Eqs.~(\ref{semi_Eqs}), we can determine the AdS$_5$ length $\ell$ by the bare negative cosmological constant $\Lambda_5$ as 
\begin{align}
\label{cosmo_const_d=5}
\Lambda_5=\frac{1}{\ell^2}\left(-6+\frac{\kappa}{\ell^2}\right) \,, \quad \kappa:=40\alpha_1+8\alpha_2+4\alpha_3.  
\end{align}
When $\kappa<0$, there is a semiclassical solution for an arbitrary negative value of $\Lambda_5$ by 
choosing a suitable length $\ell$. On the other hand, when $\kappa>0$, $\Lambda_5$ and $\kappa$ must satisfy 
\begin{align}
\label{5-dim_cosmological_constant}
\kappa|\Lambda_5|<9 \,. 
\end{align}
In the Einstein gravity limit, $\alpha_i=0$~($\kappa=0$), it reduces to the classical negative cosmological 
constant, $-6/\ell^2$.  

\section{The perturbed semiclassical static solutions}\label{sec:3}
In this section, we investigate whether or not there are perturbed static semiclassical solutions of 
Eqs.~(\ref{semi_Eqs}) with a nonzero vacuum expectation value of the stress-energy 
tensor $\Exp{\calT_{\mu\nu}}$. 
Let us expand the conformal metric $\tilde{g}_{\mu\nu}$ as 
\begin{align}
\label{series_metric}
\tilde{g}_{\mu\nu}(z,\,x)=\overline{\bdyg}_{\mu\nu}(x)+\epsilon h_{\mu\nu}(z,\,x)+O(\epsilon^2). 
\end{align} 
We assume that the metric perturbations $h_{\mu \nu}$ satisfy 
\begin{align}
\label{TT_cond}
{h_\mu}^\mu=h_{\mu\nu}\overline{\bdyg}^{\mu\nu}=0 \,, \quad
\overline{\bdyD}{}^\nu h_{\mu\nu}=0 \,,  
\end{align}
where $\overline{\bdyD}_\mu$ is the covariant derivative with respect to the unperturbed boundary metric 
$\overline{\bdyg}_{\mu\nu}$. 

Substituting the ansatz $h_{\mu\nu}(z,\,x)=\xi(z)H_{\mu\nu}(x)$ into 
the perturbation of the bulk equation~(\ref{bulk_Einstein}), one obtains the following two decoupled equations, 
\begin{align}
\label{xi_equation}
\xi''+m^2\xi-\frac{(d-1)\Omega'}{\Omega}\xi'=0, 
\end{align}
\begin{align}
\label{H_equation}
\bar{\bdyD}^2H_{\mu\nu}+\frac{2}{\ell^2}H_{\mu\nu}=m^2H_{\mu\nu}. 
\end{align} 
From the Dirichlet boundary condition at the portion of the AdS boundary $\Sigma_D$~(see Fig.~\ref{conformal_ST}), 
the solution of (\ref{xi_equation}) is analytically obtained as 
\begin{align}
\label{xi_sol}
\xi(y)=C
(1-y)^\frac{d}{2}F\left(\frac{1}{2}-p,\,\frac{1}{2}+p,\,1+\frac{d}{2};1-y  \right), 
\qquad y:=\frac{1-\cos\frac{z}{\ell}}{2}, 
\end{align}
where $p$ is defined by
\begin{align}
\label{def_p}
p=\sqrt{\frac{(d-1)^2}{4}+\hat{m}^2} 
\end{align}
with $\hat{m}^2:=m^2\ell^2$
.

As shown in the subsection~\ref{sec:3.3}, the regular static asymptotically AdS solution of 
Eq.~(\ref{H_equation}) exists only when $\hat{m}^2<0$. So, from Eq.~(\ref{def_p}), $\hat{m}^2$ must 
satisfy the inequality
\begin{align}
\label{inequality_p}
-\frac{(d-1)^2}{4}<\hat{m}^2<0
\end{align}
to have such a semiclassical solution.

\subsection{d=4 semiclassical solution}
In $d=4$ case, the solution~(\ref{xi_sol}) can be expanded near the AdS boundary, $z=0~(y=0)$ 
as~\footnote{$C$ is normalized so that $\xi(0)=1$. }
\begin{align}
\label{d=4_expan_xi}
& \xi=1+\frac{\hat{m}^2}{4}\left(\frac{z}{\ell}\right)^2+\frac{3d_1(\hat{m}^2)-\hat{m}^2}{48}\left(\frac{z}{\ell}\right)^4
+\frac{d_2(\hat{m}^2)}{16}\left(\frac{z}{\ell}\right)^4\ln \left(\frac{z^2}{4\ell^2} \right)+O(z^5), 
\nonumber \\
& d_1(\hat{m}^2)=\frac{1}{4}\left[-12-2\hat{m}^2+3\hat{m}^4
-2\hat{m}^2(2+\hat{m}^2)\left\{H\left(\frac{3}{2}-p\right)+H\left(\frac{3}{2}+p\right)\right\}\right], \nonumber \\
& d_2(\hat{m}^2)=-\hat{m}^2\left(\frac{\hat{m}^2}{2}+1  \right), 
\end{align}
where $H(k)$ is a harmonic number represented by 
the integral 
\begin{align}
H(k)=\int^1_0 \frac{1-x^k}{1-x}dx.  
\end{align}
By using the formula~(\ref{eq:general_d-variations}),
the perturbation of $\tau^{(4)}_{\mu\nu}$~(\ref{vev_SE_ct_d=4}) becomes 
\begin{align}
\label{tau_d=4}
\delta \tau^{(4)}_{\mu\nu} =
\frac{L^3}{64\pi G_5\, \ell^4}\, \left(\frac{\Hat{m}^4}{2}+\Hat{m}^2 \right)
\epsilon\, H_{\mu\nu}
\times 
\ln\frac{z^2}{4\ell^2} \,.
\end{align}
\begin{figure}[htbp]
  \begin{center}
\includegraphics[width=90mm]{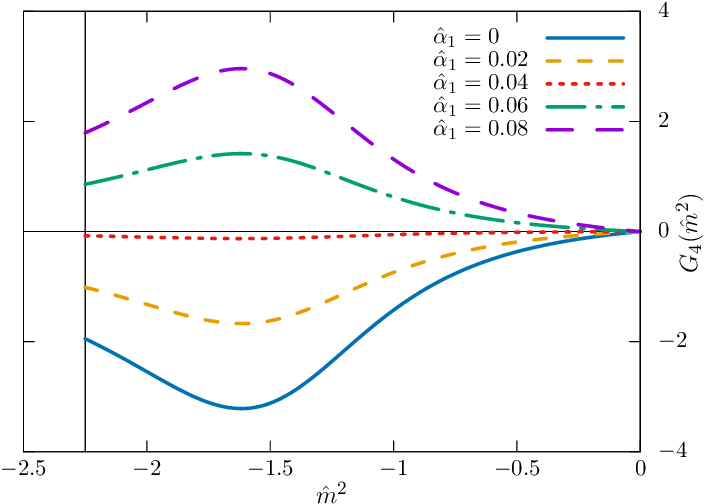}
  \end{center}
  \caption{{\small The plot of $G_4 (\hat{m}^2)$ in Eq. (\ref{4D_semi_algebraic}) for $\alpha_2=\alpha_3=0$. 
  The blue~(solid), orange~(dashed), red~(dotted), green~(dotdashed), and purple~(large dashing) curves are 
  $\hat{\alpha}_1=0, 0.02, 0.04, 0.06, 0.08$, respectively. The vertical line at $\hat{m}^2=-9/4$ represents 
  the lower bound of the mass $\hat{m}^2$. When $\hat{\alpha}_1>1/24$, there are one or two solutions, 
  depending on the value of $\gamma_4$ satisfying 
  $0<\gamma_4<G_m(\hat{\alpha}_1)$, where $G_m(\hat{\alpha}_1)$ is the maximum of 
  $G_4$ in the range of $-9/4<\hat{m}^2<0$.}}
  \label{4D_gamma4}
 \end{figure} 
Substituting Eqs.~(\ref{d=4_expan_xi}) and (\ref{tau_d=4}) into the perturbation of the stress-energy tensor (\ref{vev_SE_tensor}), the logarithmic divergence vanishes and then, we obtain a finite value as 
\begin{subequations}
\begin{align}
\label{d=4_per_SE}
  &\delta \big( \Exp{\calT_\mu{}^\nu} \big)
  = - \frac{1}{16\, \pi^, G_4}\,
  \frac{ \gamma_4\, w_4(\Hat{m}^2) }{\ell^2}\, \epsilon\, H_\mu{}^\nu,
\\
  & w_4(\hat{m}^2)
  := \frac{\pi}{8}\, \left[\,
  \left\{ H\left(\frac{3}{2}+p\right)+H\left(\frac{3}{2}-p\right) \right\}\,
  \hat{m}^2 \,\left( \hat{m}^2 + 2 \right) + 2 \Hat{m}^2 + 6
  \, \right]
  ~,
\label{eq:def-w_4}
\end{align}
\end{subequations}

On the other hand, first order perturbations of $\calE_\mu{}^\nu$ (\ref{eq:def-calE}) are given, 
with the help of Eq.~(\ref{eq:perturb-higherD-gen_Einstein_tensor-max_symm-TT-pre}),
by 
\begin{align}
  & \delta \big( \calE_\mu{}^\nu \big)
  = - \frac{ \Hat{m}^2 }{2\, \ell^2}\,
  \left\{ 1 - 6\, (4\, \Hat{\alpha}_1 + \Hat{\alpha}_2)
    + (\Hat{\alpha}_2 + 4\, \Hat{\alpha}_3)\, 
    \Hat{m}^2 
    \right\}\,
    {\epsilon}\, \pbdyg_\mu{}^\nu
  ~,
\end{align}
where $\hat{\alpha}_i:=\alpha_i/\ell^2~(i=1,2,3)$. 
the perturbed semiclassical Eqs.~(\ref{semi_Eqs})
reduce to an algebraic equation,  
\begin{align}
\label{4D_semi_algebraic}
  & \gamma_4
  = \frac{ \hat{m}^2\,
    \left\{ 1 - 6\, \left( 4\, \hat{\alpha}_1 + \hat{\alpha}_2 \right)
    + \left( \hat{\alpha}_2 + 4\, \hat{\alpha}_3 \right)\, \hat{m}^2 \right\}
    }{ w_4(\hat{m}^2) }
  =: G_4 (\hat{m}^2),
\end{align}

For simplicity, let us first consider the case $\hat{\alpha}_2=\hat{\alpha}_3=0$. The plot in 
Fig.~\ref{4D_gamma4} is the function $G_4(\hat{m}^2)$ in Eq.~(\ref{4D_semi_algebraic}) for 
$\hat{\alpha}_2=\hat{\alpha}_3=0$. 
As easily verified, there is no solution for the Einstein gravity, $\hat{\alpha}_1=0$ in the mass 
range $-9/4<\hat{m}^2<0$~(\ref{inequality_p}). 
When $\hat{\alpha}_1$ is larger than a critical value $1/24$, the function $G_4(\hat{m}^2)$ flips over, and
there is an upper bound $\gamma_{4m}$ of $\gamma_4$ to possess the solution. The upper bound 
$\gamma_{4m}$ is given by the maximum value of $G_4(\hat{m}^2)$ in the range $-9/4<\hat{m}^2<0$, and 
one or two semiclassical solutions exist, depending on the value of 
$\gamma_4<\gamma_{4m}:=G_4(\hat{m}^2)$.  
This feature is quite different from $3$-dimensional case~\cite{Ishibashi:2023luz} in which there is a lower bound 
for $\gamma_3$ to possess a solution in the range (\ref{inequality_p}). 

To describe the solution space, instead of the parameters $(\hat{\alpha}_1, \hat{\alpha}_2, \hat{\alpha}_3)$, we introduce new parameters, 
$\eta$ and $\rho$ as 
\begin{align}
\label{para_etarho}
\rho:=\frac{\hat{\alpha}_2}{2}+2\hat{\alpha}_3, \qquad 
\eta:=\frac{1}{2}-12\hat{\alpha}_1-3\hat{\alpha}_2 .
\end{align}
Since $w_4(\hat{m}^2)>0$ in the range $-9/4<\hat{m}^2<0$, the condition 
for the semiclassical solution in the mass range is summarized as 
\begin{figure}[htbp]
  \begin{center}
\includegraphics[width=110mm]{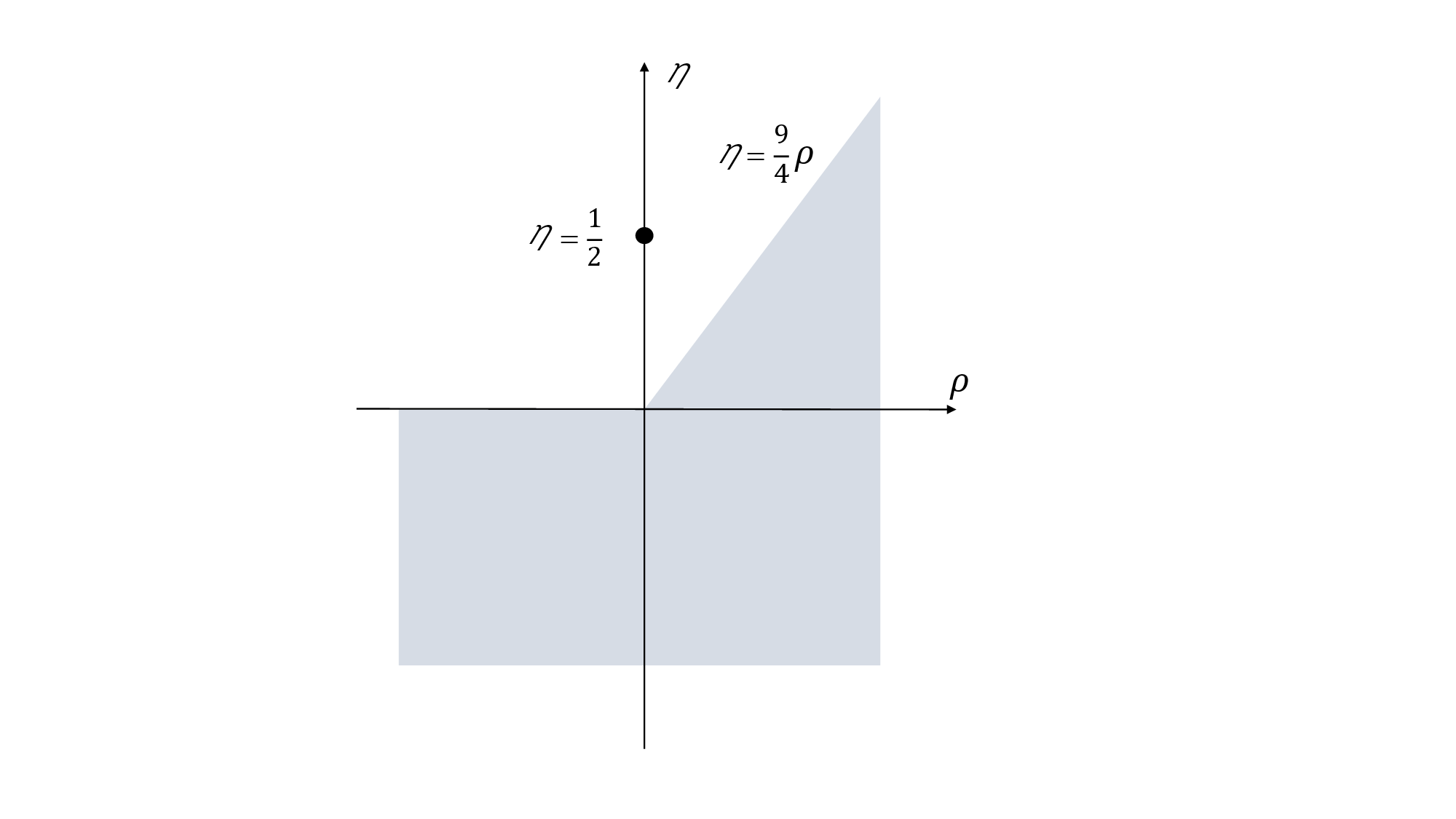}
  \end{center}
  \caption{{\small The region for the semiclassical solution to exist is shown in the shaded region. 
  The small black circle on the $\eta$ axis represents the Einstein gravity with $\hat{\alpha}_i=0$~($i=1,2,3$).}}
  \label{etarho_graph}
 \end{figure} 
\begin{align}
\label{sol4_range}
\begin{cases}
  \eta < 0 & \text{~for~~} \rho < 0
\\
  \eta < \dfrac{9}{4}\, \rho & \text{~for~~} \rho > 0 
\end{cases}
~,
\end{align}
as shown in Fig.~\ref{etarho_graph}. 

\subsection{d=5 semiclassical solution}
In $d=5$ case, the solution~(\ref{xi_sol}) can be expanded near the AdS boundary, $z=0$
as 
\begin{align}
\label{d=5_expan_xi}
\xi=1+\frac{p^2-4}{6}\left(\frac{z}{\ell}\right)^2+\frac{16-16p^2+3p^4}{72}\left(\frac{z}{\ell}\right)^4
-\frac{p(4-5p^2+p^4)\cot(\pi p)}{45}\left(\frac{z}{\ell}\right)^5+O(z^6),  
\end{align}
where $C$ is normalized as $\xi(0)=1$.
By using the formula~(\ref{eq:general_d-variations}), the perturbation 
of $\tau^{(5)}_{\mu\nu}$ is given by 
\begin{align}
\label{tau_d=5}
\delta \tau^{(5)}_{\mu\nu}
= - \frac{L^4}{72\pi G_6}\, \frac{1}{\ell^5}\, 
\left(\frac{\Hat{m}^4}{2}+\frac{9 \Hat{m}^2}{4}-9 \right)
\frac{ \epsilon\, H_{\mu\nu} }{ \sin(z/\ell) } \,.
\end{align}
At $O(\epsilon)$, the first line in Eq.~(\ref{vev_SE_tensor}) is zero, as the background solution has 
zero extrinsic curvature. Substituting Eqs.~(\ref{d=5_expan_xi}) and (\ref{tau_d=5}) into 
the second and third lines in Eq.~(\ref{vev_SE_tensor}), one obtains a finite value 
\begin{subequations}
\begin{align}
\label{per_SE5}
  & \delta ( \pExp{T_\mu{}^\nu} )
  = - \frac{1}{16\, \pi\, G_5}\, \frac{ \gamma_5\, w_5(\Hat{m}^2) }{ \ell^2}\,
  \epsilon\, H_\mu{}^\nu
  ~,
\\
  & w_5(\Hat{m}^2)
  := \frac{\pi}{9}\, \frac{ p\, (4 - p^2)\, (1 - p^2) }{ \tan(\pi p) }
  := \frac{\pi}{9}\,
  \frac{ \Hat{m}^2\, (3 + \Hat{m}^2 )\, \sqrt{ 4 + \Hat{m}^2 } }
       { \tan\left( \pi\, \sqrt{ 4 + \Hat{m}^2 } \right) }
  ~,
\label{eq:def-w_5}
\end{align}
\end{subequations}
at $O(\epsilon)$. 
From Eq.~(\ref{eq:delta-calH-all}), 
the first order perturbations of 
${\mathcal H}^{(i)}_{\mu\nu}$~(\ref{def:H}) 
is given by 
\begin{align} 
\label{dim5_deltaH}
& \delta \mathcal{H}_{\mu\nu}^{(1)}=-\frac{20}{\ell^4}
\left(2-\hat{m}^2\right)\, \epsilon\, H_{\mu\nu}, \nonumber \\
& \delta \mathcal{H}_{\mu\nu}^{(2)}=-\frac{1}{2\ell^4}
\left(\hat{m}^2-4\right)^2\, \epsilon\, H_{\mu\nu}, \nonumber \\
& \delta \mathcal{H}_{\mu\nu}^{(3)}=-\frac{2}{\ell^4}
\left(\hat{m}^4+\hat{m}^2+2\right)\, \epsilon\, H_{\mu\nu}. 
\end{align}
Also, from Eq.~(\ref{eq:perturb-higherD-gen_Einstein_tensor-max_symm-TT-pre}), 
the first order perturbations of $\calE_\mu{}^\nu$ (\ref{eq:def-calE})
are given by 
\begin{align}
  & \delta \big( \calE_\mu{}^\nu \big)
  = - \frac{ \Hat{m}^2 }{2\, \ell^2}\, \left[\, 1
    - \left\{ 10\, \left( 4\, \Hat{\alpha}_1 + \Hat{\alpha}_2 \right)
      - \Hat{\alpha}_2 \right\}
    + \left( \Hat{\alpha}_2 + 4\, \Hat{\alpha}_3 \right)\,
      \left( \Hat{m}^2 + 1 \right)
  \, \right]\,
  \epsilon\, \pbdyg_\mu{}^\nu
  ~.
\end{align}
Combining this with Eq.~(\ref{per_SE5}),
the first order perturbation of the semiclassical Eqs.~(\ref{semi_Eqs}) leads to an algebraic equation  
\begin{align}
\label{5D_semi_algebraic}
  & \gamma_5
  = \frac{ \Hat{m}^2\, \left[\, 1
    - \left\{ 10\, \left( 4\, \Hat{\alpha}_1 + \Hat{\alpha}_2 \right)
      - \hat{\alpha}_2 \right\}
    + \left( \Hat{\alpha}_2 + 4 \Hat{\alpha}_3 \right)\,  (1 + \Hat{m}^2)
    \, \right] }
    { w_5(\Hat{m}^2) }
  =: G_5(\Hat{m}^2) 
\end{align}
\begin{figure}[htbp]
  \begin{center}
\includegraphics[width=90mm]{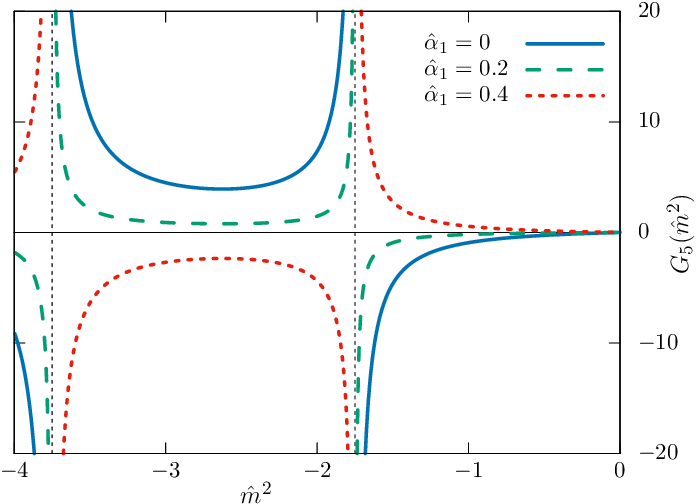}
  \end{center}
  \caption{{\small The r.h.s. of Eq. (\ref{5D_semi_algebraic}) is plotted for the Gauss-Bonnet theory 
  $\hat{\alpha}_2=-4\hat{\alpha}_1=-4\hat{\alpha}_3$.  
  The blue~(solid), green~(dashed), red~(dotted) curves are 
  $\hat{\alpha}_1=0, 0.2, 0.4$ respectively. The vertical line at $\hat{m}^2=-4$ corresponds to the 
  lower bound for the negative mass~(\ref{inequality_p}). 
  There is a lower bound $\gamma_{5c}(\hat{\alpha}_1)$ for $\gamma_5$ 
  when $\hat{\alpha}_1<1/4$ to possess the semiclassical solution. When $\hat{\alpha}_1>1/4$, 
  one solution exists in the range $-7/4<\hat{m}^2<0$ for any $\gamma_5>0$.}}
  \label{5D_gamma5}
 \end{figure} 

Let us consider, for example, the Gauss-Bonnet theory, $\hat{\alpha}_2=-4\hat{\alpha}_1=-4\hat{\alpha}_3$. 
In this case, the square bracket in the r.~h.~s.~of Eq. (\ref{5D_semi_algebraic}) becomes constant, i.~e.~, 
$1-4\hat{\alpha}_1$. Fig.~\ref{5D_gamma5} shows the plot of the r.~h.~s.~of Eq. (\ref{5D_semi_algebraic}) 
for several values of $\hat{\alpha}_1$. 
The blue~(solid), green~(dashed), red~(dotted) curves are $\hat{\alpha}_1=0, 0.2, 0.4$ respectively. 
When $\hat{\alpha}_1<1/4$, there is a lower bound $\gamma_{5c}(\hat{\alpha}_1)$ for $\gamma_5$ to 
possess the semiclassical solution. For any value of $\gamma_5>\gamma_{5c}$, there are two solutions in 
 the range $-15/4<\hat{m}^2<-7/4$. When $\hat{\alpha}_1>1/4$, the curve 
flips over~(see, for example, $\hat{\alpha}_1=0.4$ curve in Fig.~\ref{5D_gamma5}). Thus, one 
solution always exists in the range $-7/4<\hat{m}^2<0$ for any positive value of $\gamma_5$, and 
there is a possibility that there is another solution in the range $-4<\hat{m}^2<-15/4$, depending on 
$\hat{\alpha}_1$ and $\gamma_5$. 

\subsection{Static hairy AdS solutions}\label{sec:3.3}
As shown in the previous subsection, the dimensionless parameters $\gamma_4$ and $\gamma_5$ 
must satisfy Eqs.~(\ref{4D_semi_algebraic}) and (\ref{5D_semi_algebraic}) derived from the semiclassical Eqs.~(\ref{semi_Eqs}). In this subsection, we construct $d=4$ and $d=5$ dimensional static hairy global 
AdS and hyperbolic AdS solutions of Eq.~(\ref{H_equation}) under the conditions (\ref{4D_semi_algebraic}) and (\ref{5D_semi_algebraic}). We start with the $d$-dimensional metric,  
\begin{align}
\label{bdy_metric}
& ds_d^2=-\frac{f(u)}{u}(1+\epsilon T(u))dt^2+\frac{\ell^2}{4u^2f(u)}(1+\epsilon U(u))du^2
+\frac{\ell^2}{u}(1+\epsilon R(u))d\sigma_{K, d-2}^2, \nonumber \\
& f(u):=1+Ku, \qquad K=\pm 1, 
\end{align}
where $d\sigma_{K,d-2}^2=\gamma_{ij}dz^idz^j$ with $z^i$ being angle coordinates  
is the metric of the $(d-2)$-dimensional 
unit sphere or hyperbolic space, depending on $K= \pm 1$. The background solution with $\epsilon=0$ reduces to the 
global AdS~($K=1$) or the (covering space of) hyperbolic AdS~($K=-1$) black hole. From the transverse-traceless 
condition~(\ref{TT_cond}), one obtains the following two equations    
\begin{subequations}
\begin{align}
\label{traceless}
&U+T+(d-2)R=0 \,, 
\\
\label{transverse}
&u\, U' - \frac{1}{2}\, \left( d - 1 + \frac{1}{f} \right)\, U
- \frac{K}{2}\, \frac{u}{f}\, T = 0 \,. 
\end{align}
\end{subequations}
By Eq.~(\ref{H_equation}) combined with Eqs.~(\ref{traceless}) and (\ref{transverse}), one obtains 
the master equation for $U$ as 
\begin{align}
\label{static_master_U}
4u^2fU''- 2\{d+1+(d-3)Ku\}uU'+(2+2d-\hat{m}^2)U=0 
\end{align}
for $d=4,\,5$. Since $T$ and $R$ are determined by $U$ via Eqs.~(\ref{traceless}) and (\ref{transverse}), 
all metric functions are determined by the solution of Eq.~(\ref{static_master_U}). 
For $K=1$, by imposing regularity at $u=\infty$, the solution is uniquely determined as   
\begin{align}
\label{static__master_gsol}
U=F\left(\frac{3+d}{4}+\frac{p}{2},\,\frac{3+d}{4}-\frac{p}{2},\,\frac{1+d}{2};-\frac{1}{u}   \right), 
\end{align}
except the normalization factor, where $p$ is the parameter defined in Eq.~(\ref{def_p}).  
The asymptotic expansion of $U$ near the AdS boundary becomes 
\bena
\label{asy_dk=1_U}
 U&\simeq& c_1u^{\frac{3+d+2p}{4}}+c_2u^{\frac{3+d-2p}{4}} \nonumber \\
 &=&\frac{\Gamma\left(\frac{1+d}{2} \right)\Gamma(-p)}
{\Gamma\left(\frac{3+d-2p}{4} \right)\Gamma\left(\frac{d-1-2p}{4}\right)}
u^{\frac{3+d+2p}{4}}+\frac{\Gamma\left(\frac{1+d}{2} \right)\Gamma(p)}
{\Gamma\left(\frac{3+d+2p}{4} \right)\Gamma\left(\frac{d-1+2p}{4}\right)}
u^{\frac{3+d-2p}{4}}. 
\eena

For $K=-1$, we shall impose the regularity condition at the horizon $u=1$. The solution 
is given by 
\begin{align}
\label{static__master_Hysol}
U=u^{\frac{3+d-2p}{4}}F\left(\frac{3+d}{4}-\frac{p}{2},\,\frac{5-d}{4}-\frac{p}{2},\,2;1-u   \right). 
\end{align}
The asymptotic expansion of $U$ near the AdS boundary becomes 
\bena
\label{asy_dk=-1_U}
U&\simeq& c_1u^{\frac{3+d+2p}{4}}+c_2u^{\frac{3+d-2p}{4}} \nonumber \\
&=& \frac{\Gamma(-p)u^{\frac{3+d+2p}{4}}}
{\Gamma\left(\frac{5-d-2p}{4} \right)\Gamma\left(\frac{3+d-2p}{4} \right)} 
+\frac{\Gamma(p)u^{\frac{3+d-2p}{4}}}
{\Gamma\left(\frac{5-d+2p}{4} \right)\Gamma\left(\frac{3+d+2p}{4} \right)} . 
\eena

Then, from Eqs.~(\ref{traceless}) and (\ref{transverse}), we find that 
the solutions asymptotically behave as
\begin{align}
U=O\left(u^\frac{3+d-2p}{4}\right), \qquad R,\,T=O\left(u^\frac{d-1-2p}{4}\right). 
\end{align} 
By imposing the asymptotic boundary condition 
\begin{align}
\lim_{u\to 0}U,\,R,\,T=0, 
\end{align}
we find that $\hat{m}^2$ must satisfy the inequality~(\ref{inequality_p}). As shown in the previous subsection, 
the condition for the negative mass solutions are given by the semiclassical 
equations~(\ref{4D_semi_algebraic}) and (\ref{5D_semi_algebraic}). 

\section{Dynamical instability }\label{sec:4}

In this section, we study the dynamical instability of the boundary AdS spacetime by using our perturbed semiclassical Einstein equations (\ref{H_equation}). 
We first establish our stability criterion and apply it to the scalar-type perturbation case. Then, we show instability of the boundary AdS spacetime in $K=-1$ case by explicitly constructing unstable perturbation solutions to (\ref{H_equation}) in the $4$- and $5$-dimensions.    

\subsection{Stability criterion}
Let us consider the stability problem of the boundary AdS spacetimes against linear perturbations from the viewpoint of the energy integral. 
On the assumption of the time-dependence of perturbation $\propto e^{-i \omega t}$ and the separation of angular variables $z^i$, our boundary perturbation equations (\ref{H_equation}) 
can be cast into the following form:
\bena
  \hat{\omega}^2 \Phi = A \Phi \,, 
\label{def:eq:A}
\eena 
where $A$ is the second-order differential operator defined by
\bena 
 A\Phi = 2\sqrt{u}f \left\{ -\dfrac{d}{du}\left(2\sqrt{u}f \dfrac{d \Phi}{du} \right) + W(u) \Phi \right\} \,, 
\label{def:A}
\eena
with $f=1+Ku$ appeared in our boundary metric (\ref{bdy_metric}), and where $W(u)$ is the potential function depending on the dimensions and the type of perturbations, and $\Phi(u) $ a master variable constructed from $ H_{\mu \nu}$ in a certain way. We consider $A$ as an operator on the Hilbert space of square-integrable functions with respect to the inner product 
\bena
 (\Phi_1, \Phi_2) = \int \dfrac{du}{2\sqrt{u} f } \Phi_1^*(u) \Phi_2(u) \,, 
\eena
and the associated norm $ \| \Phi \| = \sqrt{(\Phi, \Phi)}$. 
It immediately follows from (\ref{def:eq:A}) that  
\bena
 \hat{\omega}^2 \| \Phi \|^2 = (\Phi, A \Phi) \,. 
\eena
This implies that if $A$ is a positive self-adjoint operator, then $\hat{\omega}$ is a real number and the equation (\ref{def:eq:A})---hence (\ref{H_equation})---does not allow mode solutions which exponentially grow in time. 

Since AdS spacetime is non-globally hyperbolic, in order to define the dynamics of perturbations in a sensible manner, one has to impose suitable boundary conditions at the AdS conformal boundary, as well as at inner boundary if exists. 
In the present case, we consider the stability under the Dirichlet boundary conditions both at the conformal infinity and inner boundary. Then, for our purpose, it suffices to examine the positivity of $A$ by taking $C^\infty_0(u)$, a space of smooth functions of compact support, as the domain of $A$. With the domain $C^\infty_0(u)$, $A$ is essentially self-adjoint, and then taking its closure makes $A$ self-adjoint with domain defined by the Dirichlet boundary conditions.  

For any $\Phi \in C^\infty_0(u)$, we find that 
\bena
(\Phi, A \Phi) = \| D \Phi \|^2 + \int du \tilde W(u)|\Phi |^2 \,, 
\label{expr:A}
\eena
where the derivative operator $D$ and the effective potential $\tilde W(u)$ are given, with some suitably chosen function $G(u)$, by~\cite{Ishibashi:2003ap,Ishibashi:2004wx} 
\bena
 D\Phi = 2\sqrt{u} f G \dfrac{d}{du} \left(\dfrac{\Phi}{G}\right) \,, \quad 
{\tilde W}(u) = W(u) - \dfrac{1}{G}\dfrac{d}{du} \left( 2\sqrt{u}f\dfrac{dG}{du}\right) \,.
\eena 
The expression (\ref{expr:A}) yields that if $\tilde W$ is positive definite for some $G$, then $A$ must also be positive definite, implying that (\ref{def:eq:A}) does not allow exponentially increasing unstable mode solutions. This is our stability criterion.

Now let us apply the above stability criteiron to linear metric perturbations on the boundary geometries considered in the previous section.  
One can in general decompose the metric perturbations $H_{\mu \nu}$ on the $d$-dimensional boundary AdS into three types called the scalar-, vector-, and tensor-type, according to their tensorial properties on the $(d-2)$-dimensional space with the metric $d\sigma_{K,d-2}^2=\gamma_{ij}dz^idz^j$ (the tensor-type exists only for $d-2 \geq3$). 
For the $K=1$ case, the stability analysis with more general boundary conditions at conformal infinity has been studied in detail~\cite{Ishibashi:2004wx}. For the $K=-1$ case, besides boundary conditions at conformal infinity, one also has to take care of inner boundary, which corresponds to the bifurcate surface of the Killing horizon with respect to $\partial/\partial t$. The master variable $\Phi$, as a function of $u$, can be constructed for each type of perturbations~\cite{Kodama:2000fa,Kodama:2003jz}. In the following we discuss whether our stability criterion is satisfied for the scalar-type perturbations. The stability analysis for the vector-type and tensor-type perturbations are given in Appendix~\ref{sec:vec:tensor}. More general, thorough analysis of cases with inner boundaries will be given elsewhere~\cite{IU}.

\subsection{Scalar-type perturbations} 

On the background, given by the metric~(\ref{bdy_metric}) with $\epsilon=0$, scalar-type perturbations can be expanded in terms of scalar harmonics ${\Bbb S}_{k}$ on the $(d-2)$-dimensional space $({\cal K}_{d-2}, d\sigma_{d-2,K}^2)$, satisfying 
$$
(\hat{\triangle} + k_{\rm S}^2){\Bbb S}=0 \,, 
$$ 
where $k_{\rm S}$ denotes the mode number along ${\cal K}_{d-2}$. For spherically symmetric case, $K=1$, $k_{\rm S}^2=l(l+d-3), \: l=0,1,2, \dots$. If we consider the vacuum Einstein gravity with a cosmological constant,   
we find that the homogeneous perturbations, i.e., those with $k_{\rm S}=0$, ${\Bbb S}= const. $, correspond to adding perturbatively a mass term to the background AdS metric, hence are not dynamical ones. However, in the present case, as a holographic semiclassical problem, we are dealing with the non-vacuum boundary Einstein equations with CFT source terms, 
the homogeneous perturbations can be dynamical. In other words, we are dealing with effectively the scalar-type perturbations of a massive tensor field with the mass-squared $\hat{m}^2$ on the $d$-dimensional AdS background, for which the exceptional mode $k_{\rm S}=0$ can be dynamical. As examined in detail in~\cite{Cardoso:2019mes}, for massive tensor fields, it does not, in general, appear to be possible to find a set of decoupled master equations for generic modes of scalar-type perturbations. In what follows we focus only on the exceptional mode for the scalar-type perturbations.

The scalar-type homogeneous perturbations ($k_{\rm S}=0$) in the background of general dimension $d\geq 4$ with $f=1+ Ku $ can be expressed in terms of 4 functions $R, S, T, U$ of $u$ in the following way:
\bena
\label{bdy_TDmetric}
 ds_d^2 &=&-\frac{f}{u}(1+\epsilon T(u)e^{-i\omega t})dt^2+\frac{\ell^2}{4u^2f}(1+\epsilon U(u)e^{-i\omega t})du^2
\nonumber \\
&{}& +\frac{\epsilon \ell}{2u^2f}S(u)e^{-i\omega t}dtdu
+\frac{\ell^2}{u}(1+\epsilon R(u)e^{-i\omega t})d\sigma_{K,(d-2)}^2 \,. 
\eena
The traceless and transverse 
conditions~(\ref{traceless}), (\ref{transverse}) are written by  
\begin{subequations}
\begin{align}
\label{tracelss_dyn}
& (d-2)R+T+U=0, \\
\label{transverse_dyn1}
& S'-i\hat{\omega}T-\frac{d+1}{2u}S=0, \\
\label{transverse_dyn2}
& uU'-\frac{1}{2}\left(\frac{1}{f}+d-2\right)U+\frac{T}{2f}+\frac{i\hat{\omega}}{4f^2}S+\frac{d-2}{2}R=0, 
\end{align}
\end{subequations}
where $\hat{\omega}=\ell \omega$. 
Eliminating $R$ by Eq.~(\ref{tracelss_dyn}), we also obtain the following constraint equation from 
(\ref{H_equation}) as 
\begin{align}
\label{constr_dyn}
i(\hat{m}^2-2\hat{\omega}^2)S+2\hat{\omega}u(1+\hat{\omega}^2)T+\{1+\hat{\omega}^2u-(d-1)f\}U=0. 
\end{align}
This equation yields two coupled first order differential equations for $(U, S)$ from 
Eqs.~(\ref{transverse_dyn1}) and (\ref{transverse_dyn2}). Then, defining a new variable $Z$ as
\begin{align}
\label{def_Z}
Z=S+2i\hat{\omega}fU, 
\end{align}
we find the equation for the master variable $Z$ as    
\bena
  Z'' + \left( \dfrac{f'}{f}- \dfrac{d+1}{2u}\right) Z' + \dfrac{1}{4u^2f^2} \{\hat{\omega}^2 u + [2(d+1)-\hat{m}^2]f \}Z=0 \,. 
\label{def:eq:scalar}
\eena

To examine our stability criterion, we introduce the new variable
\bena
 Z(u) = u^{(d+2)/4} \Phi(u) \,. 
\eena
Then, the equation (\ref{def:eq:scalar}) for $Z$ is rewritten as the equation for $\Phi$ in the form of (\ref{def:eq:A}) with 
\bena
 W(u) 
 &=& \frac{1}{2\sqrt{u}} \left\{ \left[ p^2 - \dfrac{(d+3)^2}{4} \right] \dfrac{1}{u} -(d+2)f' + \dfrac{(d+2)(d+4)}{4}\dfrac{f}{u} \right\} \,, 
\label{def:W}
\eena
where the parameter $p = (1/2)\sqrt{(d-1)^2+ 4 \hat{m}^2}$ is introduced in (\ref{def_p}). 

\subsubsection{Global chart $K = 1$}
In this case,  $u=0$ corresponds to the AdS conformal boundary and $u=\infty$ to the regular center of the AdS spacetime. Choosing our $G$ as
\bena
\label{K+G}
 G = u^{(1-2p)/4} f^{ -(d+3-2p)/4}
\eena
we find (\ref{expr:A}) with 
\bena
 \tilde{W}(u)= \frac{1}{8\sqrt{u} f} \left( d+3 - 2p \right)^2 \geq 0\,. 
\eena
Thus, for any $\Phi \in C^\infty_0(u)$ with $0 < u < \infty$, 
\bena
  (\Phi, A \Phi) = \| D \Phi \|^2 + \frac{1}{4} \left( d+3 - 2p \right)^2 \|\Phi \|^2 \geq 0 \,.
\eena
Therefore the equation (\ref{def:eq:A}) does not allow exponentially growing unstable modes for homogeneous perturbations. 
We should note that $\Phi$ describes only a highly restricted class of perturbations---$(d-2)$-dimensional spherically symmetric perturbations, and 
the above analysis does not assure, at this stage, the stability of AdS in global chart against linear perturbations.

\subsubsection{Hyperbolic chart $K = -1$}

In this case, $u=0$ corresponds to the AdS conformal boundary and $u=1$ to the bifurcate surface of the Killing horizon with respect to the Killing field $\partial /\partial t$, where for the stability analysis, we impose the Dirichlet boundary condition\footnote{  
Note that instead of the Dirichlet boundary condition, one may impose the ingoing boundary conditions at the horizon $u=1$, and discuss the stability in terms of quasi-normal modes, rather than the energy integral (\ref{expr:A}) employed in this section. (Note that for perturbations $\Phi$ satisfying the ingoing boundary conditions, the norm $\|\Phi\|$ will be divergent.) One may wonder what happens for perturbations which do not satisfy the Dirichlet conditions at the bifurcate surface. It is shown that if the Schwarzschild wedge of the spacetime is stable under the Dirichlet boundary conditions at the bifurcate surface, then by using the discrete isometry of the Kruskal extended spacetime, one can rule out the possibility that initial data which are non-vanishing at the bifurcate surface could exhibit instability~\cite{Kay:1987ax}. This argument applies to the present case of hyperbolic AdS ($K=-1$) with the bifurcate Killing horizon.   
}. 

For $K=-1$, the choice (\ref{K+G}) does not work, making $\tilde W$ negative definite. 
Now, by inspection, we choose our $G$ as
\bena
 G =  u^{\left(1 +2p \right)/4} \,, 
\eena
then 
\bena
 \tilde W = \dfrac{1}{2\sqrt{u}} \left\{ (p +1)^2 - \dfrac{(d+1)^2}{4} \right\} \,.  
\eena
This effective potential $\tilde W(u)$ can only be non-negative when $p \geq (d-1)/{2}$ (i.e., when $\hat{m}^2 \geq 0$). This leaves open the possibility of the existence of unstable modes for $p < (d-1)/2$, evading our stability criterion. We will see that this is indeed the case for $4$- and $5$-dimensions in the next subsection.

\subsection{Unstable solutions}\label{sbsc:unstable} 
We show that the $4$ and $5$-dimensional background hyperbolic AdS solutions are unstable against perturbation satisfying the semiclassical equations~(\ref{4D_semi_algebraic}) 
and (\ref{5D_semi_algebraic}). 
The general solution to the master equation (\ref{def:eq:scalar}) with $K=-1$ is given by 
\begin{align}
\label{Z_sol_general}
   Z
  &= (1-u)^{\frac{-i\hat{\omega}}{2}}\, u^{\frac{3+d}{4}}\,
  \left\{
  a_1\, u^{-\frac{p}{2}}\,
     F\left( - \frac{ d-1+2p+2i\hat{\omega} }{4}, \,
             \frac{ 3+d-2p-2i\hat{\omega} }{4},\,1-p\,;\,u  \right)  
\right.
\nonumber \\
  &\hspace*{3.5truecm}
\left.
  + a_2\, u^{ \frac{p}{2} }\,
  F\left( - \frac{ d-1-2p+2i\hat{\omega} }{4}, \,
    \frac{ 3+d+2p-2i\hat{\omega} }{4},\,1+p\,;\,u  \right)
  \right\}
  ~.
\end{align}

We impose the Dirichlet boundary condition at the AdS conformal boundary at $u=0$. This corresponds to $a_1=0$. As for the inner boundary, since we are now constructing an unstable solution rather than showing the stability, instead of the Dirichlet conditions according to \cite{Kay:1987ax}, we impose the ingoing boundary condition at the Killing horizon, which corresponds to 
$Z\sim (1-u)^{-\frac{i\hat{\omega}}{2}}$. By the transformation formula~(\ref{Hyper_trans}), the solution near the horizon 
behaves as 
\bena
\label{Z_sol_nearH}
Z &\simeq &
\frac{ \Gamma(-i\hat{\omega})\Gamma(1+p)
(1-u)^{  i\Hat{\omega}/2  }
}
{\Gamma\left(-\frac{1}{4}(d-1-2p+2i\hat{\omega})\right)
\Gamma\left(\frac{1}{4}(3+d + 2\, p -2i\hat{\omega})\right)}
\nonumber \\
&+&\frac{\Gamma(i\hat{\omega})\Gamma(1+p)
(1-u)^{ - i \Hat{\omega}/2  } }
{
\Gamma\left(-\frac{1}{4}(d-1-2p-2i\hat{\omega})\right)
\Gamma\left(\frac{1}{4}(d+3 + 2\, p +2i\hat{\omega})\right)
}. 
\eena
Thus, for an arbitrary non-negative integer $n$, the ingoing boundary condition requires that 
\begin{align}
\hat{\omega}=i\left(\frac{d-1}{2}-2n-p   \right)\quad \text{or} \quad \hat{\omega}=-i\left(\frac{d+3}{2} +2n+p   \right). 
\end{align} 
The modes that satisfy the left-side condition above with $n=0$ correspond to unstable modes, since $p<(d-1)/2$ for the 
negative mass~(see Eq.~(\ref{def_p})).  
This implies that the hyperbolic AdS solution is unstable against the semiclassical perturbation. 

\section{The boundary free energy}\label{sec:5}
In this section, we investigate thermodynamical instabilities of the static hairy solutions obtained in 
Sec.~\ref{sec:3} by evaluating the free energy. In general theory of gravity with quadratic curvature 
corrections~(\ref{bd_action}), some ambiguity arises for evaluating the free energy since it includes 
higher derivatives of the metric functions. This implies that the surface term arising from the on-shell 
action~(\ref{bd_action}) includes higher derivatives of the boundary metric, where there is no natural guiding  
principle to choose the values of the derivatives. In this paper, we restrict our attention to the Gauss-Bonnet 
gravity satisfying the parameter relation  
\begin{align}
\label{Coeff_GB}
-\frac{\hat{\alpha}_2}{4}=\hat{\alpha}_3=\hat{\alpha}_1. 
\end{align}
In this case, the ambiguity mentioned above does not arise, as the surface term of the action~(\ref{bd_action}) 
does not include the higher derivatives. 

In the $4$-dimensional case, it is well known that the Gauss-Bonnet theory does not 
affect the equations of motion~(\ref{semi_Eqs}), and that it is the same as the one of the Einstein 
gravity with $\hat{\alpha}_1=0$. Actually, one can check that the parameters $\rho$ and $\eta$ in 
(\ref{para_etarho}) reduce to the Einstein gravity with $(\rho, \,\eta)=(0,\,1/2)$. As shown in 
Fig.~\ref{etarho_graph}, there is no hairy global AdS~(hyperbolic AdS BH) solution in the Einstein gravity.   
So, hereafter, we shall calculate the free energy in the $5$-dimensional Gauss-Bonnet gravity.  

The generalized Gibbons-Hawking term ${\mathcal S}_\text{GH}$ and the counter term ${\mathcal S}_\text{ct}$ 
in the Gauss-Bonnet action~(\ref{bd_action}) with the relation~(\ref{Coeff_GB}) 
are given by 
\bena
\label{bd_surface_term}
 {\cal S}_\text{GH} &=&\frac{1}{ 8 \pi G_5}\int d^4x\sqrt{-\sigma}
\Biggl[\bdyK+2\alpha_1\Bigl(
                                      \bdyK\bdyR^{(4)}-2{\bdyK_a}^b{\bdyR_b}^a +\bdyK{\bdyK_a}^b{\bdyK_b}^a 
                                      \nonumber \\
&{}& \qquad \qquad \qquad \qquad 
                                     -\frac{2}{3}{\bdyK_a}^b{\bdyK_b}^c{\bdyK_c}^a-\frac{1}{3}\bdyK^3
                              \Bigr)          
                                     \Biggr] \Biggl{|}_{u=0} \,, 
\nonumber \\
 {\cal S}_\text{ct} &=& \frac{3}{16\pi G_5}\int d^4x\sqrt{-\sigma}
                             \left( 
                                    \frac{\alpha}{\ell}-\beta\ell {\cal R}^{(4)}  
                             \right) \Biggl{|}_{u=0} \,,  
\eena
where $\sigma_{ab}:=\bdyg_{ab}$ and $\bdyK_{ab}$ is the $4$-dimensional extrinsic curvature given by 
\begin{align}
{\mathcal K}_{ab}:=-\frac{ \sqrt{g^{uu}}}{2}\p_u\sigma_{ab}=-\frac{u\sqrt{f}}{\ell\sqrt{1+\epsilon U}}\p_u \bdyg_{ab}, 
\end{align}   
and $\alpha$ and $\beta$ are the free parameters determined below.    

The action~(\ref{bd_action}) combined with (\ref{bd_surface_term}) can be expanded as a series in
$\epsilon$ as 
\begin{align}
\label{def_expansion_action}
{\mathcal S}=\sum_{n=0}^\infty \epsilon^n\,\delta^{(n)}{\mathcal S}. 
\end{align}  
First, let us consider the first order variation of the action~(\ref{bd_action}) with (\ref{bd_surface_term}).  
Under the traceless and transverse condition~(\ref{TT_cond}), the variation of ${\mathcal S}_\text{bd}+S_\text{bulk}$ 
is zero under the on-shell condition~(\ref{semi_Eqs}), and the variations of the other terms are
\begin{align}
\label{first_variation}
& \delta^{(1)}{\mathcal S}_\text{GH}
\simeq -\frac{c_2\ell^2}{16\pi G_5}\int_{\p\Sigma_0}dtd{\mathcal V}_K \,
\left(4+p+4p\hat{\alpha}_1 \right)u^{-\frac{p}{2}}, \nonumber \\
& \delta^{(1)}{\mathcal S}_\text{ct}\simeq \frac{3c_2\ell^2}{16\pi G_5}
\int_{\p\Sigma_0}dtd{\mathcal V}_K
\,\left\{-\frac{\alpha}{2}+2\beta(1+p) \right\}u^{-\frac{p}{2}}, 
\end{align}
where $d{\mathcal V}_K$ is the volume element of the $3$-dimensional metric $d\sigma^2_{K, 3}$ 
in (\ref{bdy_metric}). We choose the parameters $\alpha$ and $\beta$ so that the total divergence of 
$\delta^{(1)}{\mathcal S}_\text{GH}+\delta^{(1)}{\mathcal S}_\text{ct}$ vanishes. The condition is 
\begin{align}
\label{sousatu1}
-\frac{3\alpha}{2}+6(1+p)\beta=p+4+4p\hat{\alpha}_1. 
\end{align}
Substituting the perturbed metric~(\ref{bdy_metric}) into
Eq.~(\ref{second_variation_bd_action}),  
and using Eqs.~(\ref{traceless}), (\ref{transverse}), and (\ref{static_master_U}), we
obtain 
\bena
\label{res_second_bd}
\delta^{(2)} {\mathcal S}_\text{bd} + \delta^{(2)} S_\text{bulk}
&=&\frac{c_2^2\ell^2(1+p)^2}{48\pi G_5}u^{-p}
\int_{\p\Sigma_0}dtd{\cal V}_K \, \{3p-4+4\hat{\alpha}_1(8-3p-2p^2)  \}   
\nonumber \\
  &{}& -\frac{c_1c_2\ell^2}{6\pi G_5}(1-p^2)\int_{\p\Sigma_0}dtd{\cal V}_K \,
\{1-2(p^2+4)\hat{\alpha}_1   \}+O(u^{1-p}), 
\eena
where we used Eqs.~(\ref{traceless}), (\ref{transverse}), (\ref{static_master_U}), (\ref{asy_dk=1_U}), 
and (\ref{asy_dk=-1_U}). Similarly, expanding the surface terms~(\ref{bd_surface_term}) with respect to 
$\epsilon$, up to $O(\epsilon^2)$, we obtain 
\bena
\label{second_surface_term}
 \delta^{(2)}{\mathcal S}_\text{GH} &=& -\frac{ c_2^2\ell^2(1+p)^2}{12\pi G_5}u^{-p}
\int_{\p\Sigma_0}dtd{\cal V}_K \,
\{p-2(p^2+2p-4)\hat{\alpha}_1  \}   
\nonumber \\
&{}& +\frac{c_1c_2\ell^2}{3\pi G_5}\int_{\p\Sigma_0}dtd{\cal V}_\pm\,
(p^4+3p^2-4)\hat{\alpha}_1+O(u^{1-p}) \,, 
\nonumber \\
\delta^{(2)}{\mathcal S}_\text{ct} &=&-\frac{\ell^2\alpha}{16\pi G_5}
\int_{\p\Sigma_0}dtd{\cal V}_\pm\left[c_2^2(1+p)^2u^{-p}+2c_1c_2(1-p^2)\right]+O(u^{1-p}). 
\eena
It is easily checked that the total leading divergent terms 
$\delta^{(2)}{\mathcal S}$ at $O(u^{-p})$ vanishes when 
\begin{align}
\label{sousatu2}
-8p\hat{\alpha}_1-8-2p-3\alpha+12(1+p)\beta=0 
\end{align}
is satisfied. By solving Eqs.~(\ref{sousatu1}) and (\ref{sousatu2}), the free parameters $\alpha$ and $\beta$ are
uniquely determined as 
\begin{align}
\label{sol_alpha_beta}
\alpha=-\frac{4+p}{3}+\frac{4}{3} p\,\hat{\alpha}_1, \qquad \beta=\frac{4+p+12p\hat{\alpha}_1}{12(1+p)}. 
\end{align}
For the Einstein gravity~($\hat{\alpha}_1=0$) with $\hat{m}^2=0$, the parameter reduces to 
$\alpha=-2$ and $\beta=1/6$, which is the same as the counter term in the $5$-dimensional 
AdS bulk gravity theory~\cite{Balasubramanian:1999re}. 

When $p>1$ one cannot obtain a finite value in the second variation $\delta^{(2)}{\mathcal S}$, although 
the leading divergent term cancels each other because the subleading terms also diverge at 
$O(u^{1-p})$, as seen in Eqs.~(\ref{res_second_bd}) and (\ref{second_surface_term}). 
So, we shall pay attention to the case $p<1$ where the subleading term converges to zero. In this case, 
the total second order variation yields the following finite term, 
\bena
\label{second_order_ft} 
 \delta^{(2)}{\mathcal S} &=&
\delta^{(2)} {\mathcal S}_\text{bd} + \delta^{(2)} S_\text{bulk}
 +\delta^{(2)}{\mathcal S}_\text{GH}+\delta^{(2)}{\mathcal S}_\text{ct}
\nonumber \\
&=&\frac{\ell^2c_1c_2\,p(1-p^2)(1-4\hat{\alpha}_1)}{24\pi G_5}\int_{\p\Sigma_0}dtd{\cal V}_\pm. 
\eena
For the asymptotically global AdS case, $K=1$, by Eq.~(\ref{asy_dk=1_U}), one obtains 
\begin{align}
\label{c1c2_global}
c_1c_2=-\frac{32\tan\left(\frac{p\pi}{2} \right)}{\pi p^3(4-p^2)}<0, \,\,\text{for}\,\, 0<p<1,  
\end{align}
and for the hyperbolic AdS BH case, $K=-1$, by Eq.~(\ref{asy_dk=-1_U}), one obtains 
\begin{align}
\label{c1c2_BH}
c_1c_2=\frac{2\tan\left(\frac{p\pi}{2} \right)}{\pi p(4-p^2)}>0,  \,\,\text{for}\,\, 0<p<1. 
\end{align}

The deviation $\Delta F$ of the free energy of our static semiclassical solutions constructed in section~\ref{sec:3}
from that of the corresponding (either global AdS or hyperbolic AdS BH) background is related to the total 
effective action by 
\bena
\label{free energy}
 \Delta F &=&
-({\mathcal S}^\text{OS}-\bar{{\mathcal S}}^\text{OS})\Bigr{/}\int dt \nonumber \\
 &=&-\frac{\epsilon^2}{24\pi G_5}\ell^2c_1c_2\,p(1-p^2)(1-4\hat{\alpha}_1)\int d{\cal V}_\pm+O(\epsilon^3). 
\eena
By Eqs.~(\ref{c1c2_global}) and (\ref{c1c2_BH}), the hairy hyperbolic AdS black hole solution~(\ref{static__master_Hysol})  
is thermodynamically stable when $\alpha_1<1/4$. On the other hand, for $\alpha_1>1/4$, another hairy 
solution appears in the range $0<p<1$~($-4<\hat{m}^2<-3$) when 
$\gamma_5$ is larger than a critical value, as shown in Fig.~\ref{5D_gamma5}. In this case, the hairy 
global AdS solution~(\ref{static__master_gsol}) becomes thermodynamically stable.

\section{Summary and discussions}\label{sec:6}
We have investigated $4$ and $5$-dimensional semiclassical asymptotically AdS solutions in the quadratic theory of gravity 
in the context of the AdS/CFT duality. The solutions are perturbatively constructed from the maximally 
symmetric AdS background solutions. One type of the solutions is an asymptotically global AdS solution with no horizon, and 
the other is the asymptotically hyperbolic AdS black hole solution with a Killing horizon.  

The presence of such a solution is determined by the parameters $\hat{\alpha}_i:=\alpha_i/\ell^2~(i=1,2,3)$ in 
the quadratic gravity action~(\ref{bd_action}), where $\ell$ is the boundary AdS length. 
In the $4$-dimensional case, the solutions with non-trivial quantum hair exist only in the 
parameter region~(\ref{sol4_range}) where the Einstein gravity with $\hat{\alpha}_i=0$ is excluded. 
In the $5$-dimensional case, the parameter space for the existence of such a solution is 
complicated. In the Gauss-Bonnet theory, for example, there is a lower bound for the gravitational constant $G_5$ to 
possess such a hairy solution when $\hat{\alpha}_1<1/4$~(including the Einstein gravity), while the solution always 
exists when $\hat{\alpha}_1>1/4$~(not including the Einstein gravity), as shown in Sec.~\ref{sec:3}. 
In both the $4$ and $5$-dimensional hyperbolic AdS black hole cases, the background maximally 
symmetric (zero mass) hyperbolic AdS solutions are dynamically unstable when the corresponding hairy AdS black hole solutions exist, as shown in Sec.~\ref{sec:4}. 

When the parameters $\hat{\alpha}_i$ are small enough,  i.e.,~$|\hat{\alpha}_i|\ll 1$, the existence of 
such a solution crucially depends on the dimension $d$ of the boundary spacetime. 
Let us consider a null vector $l^\mu=\bar{l}^\mu+\delta l^\mu$ satisfying $\bdyg_{\mu\nu}l^\mu l^\nu=0$, 
where $\bar{l}^\mu$ is the null vector on the background metric $\overline{\bdyg}_{\mu\nu}$ and 
$\delta l^\mu$ is the first order deviation. Contracting 
the semiclassical Eqs.~(\ref{semi_Eqs}) with the null vector $l^\mu$ and taking the first order perturbation of 
it by using Eq.(\ref{eq:general_d-variations}), we obtain 
\begin{align}
\label{linear_response}
  & 2\, \delta\big( \bdyR_{\mu\nu}\, l^\mu\,  l^\nu \big)
  = - m^2\, \delta\bdyg_{\mu\nu}\, \Bar{l}^\mu\, \Bar{l}^\nu 
\nonumber \\
  &\hspace*{2.4truecm}
  \simeq 16\, \pi\, G_d\,
    \delta\big( \Exp{\calT_{\mu\nu}}\, l^\mu\, l^\nu \big)
  = - \dfrac{\gamma_d\, w_d}{\ell^2}
  \delta\bdyg_{\mu\nu}\, \Bar{l}^\mu\, \Bar{l}^\nu \,, 
\end{align}
and thus 
\begin{align} 
\Hat{m}^2 \simeq \gamma_d\, w_d \,, 
\end{align} 
where $w_d$ is defined
as
\begin{align}
   \delta\big( \Exp{\calT_\mu{}^\nu} \big)
  &= \delta\big( \Exp{ \calT_{\mu\alpha} } \big)\,
    \overline{\bdyg}^{\alpha\nu}
  - \overline{ \Exp{ \calT_{\mu\alpha} } }\, \delta\bdyg^{\alpha\nu}
\nonumber \\
  &=: - \frac{1}{16\, \pi\, G_d}\, \frac{ \gamma_d\, w_d }{\ell^2}\,
    \delta\bdyg_\mu{}^\nu
  ~,
\end{align}
and in the equality of the second line in Eq.~(\ref{linear_response}), we used the fact that $\overline{\Exp{\calT_{\mu\nu}}}$ is proportional to 
$\overline{\bdyg}_{\mu\nu}$ if it takes a non-zero value~(\ref{vev_d=4_bg_SE}).  

For the cases of $d = 4, 5$, the coefficient $w_d$ is given respectively by Eqs.~(\ref{eq:def-w_4}) and (\ref{eq:def-w_5}). 
As shown in Sec.~\ref{sec:3}, the mass $m^2$ should be negative
to possess an asymptotically AdS static solution,
implying that the coefficient $w_d$ must be {negative}.
It is easily checked that $w_4$ is always {positive} 
in the whole region $-9/4<\hat{m}^2<0$ in Eq.~(\ref{inequality_p}),
while $w_5$ can be {negative} in the range $-15/4<\hat{m}^2<-7/4$ in the region
satisfying (\ref{inequality_p}).
So, there is no semiclassical solution in $d=4$ case, while there are two solutions in the same parameter range in $d=5$ case when 
the parameters $\hat{\alpha}_i$ are small enough. Since the change of sign in $w_5$ is similar to the 
$d=3$ case~\cite{Ishibashi:2023luz}, we expect that it is a peculiar property of the odd dimension.    

From the perspective of the linear response theory, $w_d$ is nothing but the response function of the 
metric perturbation $\bdyg_{\mu\nu}$ and it depends on the state of the boundary field theory. According to the 
AdS/CFT dictionary, the boundary state is determined by the bulk structure such as a bulk black hole representing 
a thermal state or AdS bubble representing a confining phase.  
In our case, $w_d$ would depend on the boundary condition at another AdS boundary $\Sigma_D$ via 
Eq.~(\ref{xi_equation}). So, it would be interesting to explore the case of different boundary conditions apart from the 
Dirichlet boundary condition or to extend our analysis into the framework of the AdS/BCFT 
duality~\cite{Takayanagi:2011zk} by inserting some end of the world branes in the bulk.   
   
In section~\ref{sec:5}, we have shown in the $5$-dimensional Gauss-Bonnet theory that 
the free energy of the asymptotically hyperbolic AdS BH solution is smaller than that of the maximally 
symmetric hyperbolic AdS BH solution, provided that $\hat{\alpha}_1<1/4$. 
Since the solution has a bifurcate (non-degenerate) Killing horizon, the boundary state corresponds to a thermal state, implying 
that the hairy solution is thermodynamically more favorable than that of the maximally symmetric hyperbolic 
AdS BH solution. It would be straightforward to extend our analysis to the $4$ and $5$-dimensional 
Schwarzschild AdS black holes or Kerr-AdS black holes. Since asymptotically AdS spacetime acts like 
a confining box, one expects that there are semiclassical AdS black hole solutions with quantum hair, 
representing the Hartle-Hawking-like equilibrium states. However, the present analysis seems to exclude 
the existence of such a semiclassical AdS black hole solution in the $4$ dimensional case (at least in the 
Einstein gravity limit), since $w_4$ is always positive, unless the semiclassical solution with $m^2>0$ does not exist. 
The complete analysis would appear in the near future.

\begin{center}
{\bf Acknowledgments}
\end{center}
We wish to thank Masahiro Hotta {and Makoto Natsuume} for useful discussions. 
We are grateful to the long term workshop YITP-T-23-01 held at YITP,
Kyoto University, where a part of this work was done. 
This work is supported in part by JSPS KAKENHI Grant No. 15K05092, 20K03938 (A.I.), 20K03975 (K.M.) and also supported by MEXT KAKENHI Grant-in-Aid for Transformative Research Areas A Extreme Universe No.21H05186 (A.I. and K.M.) and 21H05182. 

\appendix

\section{Variation formulas}
When linear metric perturbations $\delta\bdyg_{\mu\nu} =: \epsilon\, H_{\mu\nu}$ of the $d$-dimensional AdS background satisfy the transverse-traceless 
conditions 
($\overline{\bdyg}{}^{\mu\nu} H_{\mu\nu} = \BGbdyD_\nu H_\mu{}^\nu = 0$),
and also Eq.~(\ref{H_equation}), the first order perturbations of
the curvature tensor and its derivatives are given as follows: %
%
%
\begin{subequations}
\label{eq:general_d-variations}
\begin{align}
  & \delta \left( \bdyR^{\mu\nu}_{\rho\sigma} \right)
  = \epsilon\, \left( - 2\,
  \BGbdyD{}^{[\mu} \BGbdyD_{[\rho}
    \pbdyg_{\sigma]}^{\nu]}
  + \frac{2}{\ell^2}\, \delta_{[\rho}^{[\mu}\, \pbdyg^{\nu]}_{\sigma]}
  \right)
  ~,
\label{A-eq:vary-Riemann^KL_MN-max_symm_bg-II} \\
  & \delta\, \left( \bdyR_{\mu}^{\nu} \right)
  = - \frac{m^2}{2}\, \epsilon\, \pbdyg_{\mu}^{\nu}
  ~,
& & \delta \bdyR = 0
  ~,
\intertext{and, }
  & \delta \left( \bdyD_\rho \bdyR_{\mu}^{\nu} \right)
  = - \epsilon\, \frac{m^2}{2}\,
  \BGbdyD_\rho
  \pbdyg_{\mu}^{\nu}
  ~,
& & \delta\, \left( \bdyD^2 \bdyR_{\mu}^{\nu} \right)
  = - \epsilon\, \frac{m^2}{2}\,
  \BGbdyD^2
  \pbdyg_{\mu}^{\nu}
  ~,
\label{eq:vary-square-Ricci_tensor} \\
  & \delta\, \left( \bdyD_{\mu} \bdyD^{\nu} \bdyR \right)
  = 0
  ~,
& & \delta\, \left( \bdyD^2 \bdyR \right)
  = 0
  ~.
\label{eq:vary-nabla_nabla-Ricci_scalar}
\end{align}
\end{subequations}
%
%
From the above formulas, we have %
\begin{subequations}
\label{eq:delta-calH-all}
\begin{align}
  & \delta \big( \calH^{(1)}_\mu{}^\nu \big)
  = \frac{ d\, (d - 1) }{\ell^2}\, m^2\,
    \epsilon\, \pbdyg^{\nu}_{\mu}
  ~,
\\
  & \delta \big( \calH^{(2)}_\mu{}^\nu \big)
  = - \frac{m^2}{2}\, \left( m^2 - 2\, \frac{d - 1}{\ell^2} \right)\,
    \epsilon\, \pbdyg^{\nu}_{\mu}
  ~,
\\
  & \delta \big( \calH^{(3)}_\mu{}^\nu \big)
  = - 2\, m^2\, \left( m^2 + \frac{d - 4}{\ell^2} \right)\,
  \epsilon\, \pbdyg^{\nu}_{\mu}
  ~.
\end{align}
\end{subequations}
Then we obtain the first order perturbations of $\calE_\mu{}^\nu$ (\ref{eq:def-calE}) as 
%
\begin{align}
  & \delta \big( \calE_{\mu}^{\nu} \big)
  = - \frac{m^2}{2}\,
  \left\{ 1 - \frac{ 2\, (d - 1)\, (d \times \alpha_1 + \alpha_2)
    - 4\, (d - 4)\, \alpha_3 }{\ell^2}
    + (\alpha_2 + 4\, \alpha_3)\, m^2 \right\}\, \epsilon\, \pbdyg_\mu{}^\nu
  ~.
\label{eq:perturb-higherD-gen_Einstein_tensor-max_symm-TT-pre}
\end{align}
%
\section{Stability with respect to the vector- and tensro-type perturbations}\label{sec:vec:tensor}

We discuss the stability against the vector and tensor-type perturbations. 
In the following, $z^i$ with indices $i,j,k,\dots$ denotes angular coordinates in $(d-2)$-dimensional space ${\cal K}_{d-2}$, the unit sphere $K=1$ or hyperbolic space $K=-1$, and the coordinates $y^a=(t,u)$ with indices $a,b,\dots$ denotes the $2$-dimensional part of $g_{\mu \nu}$. 

\subsection{Vector-type perturbations} 
The vector-type perturbations can be given by 
\bena
  H_{ab}=0 \,, \quad H_{ai} = \dfrac{\ell}{\sqrt{u}}  f_a {\Bbb V}_i \,,  \quad H_{ij} =2\dfrac{\ell^2}{u}H_T{\Bbb V}_{ij} \,, 
\eena
where ${\Bbb V}_i$ is the transverse vector harmonics on the space ${\cal K}_{d-2}$, satifying 
\bena (\hat{\triangle} + k_{\rm V}^2){\Bbb V}_{i}=0 \,, \quad \hat{D}^i{\Bbb V}_i=0 \,, 
\eena 
with $\hat{D}_i$ the derivative operator compatible with $\gamma_{ij}$ and $\hat{\triangle} = \gamma^{ij} \hat{D}_i \hat{D}_j$. 
Here, 
\bena
 {\Bbb V}_{ij} = - \dfrac{1}{2k_{\rm V}}(\hat{D}_i{\Bbb V}_{j}  + \hat{D}_j{\Bbb V}_{i}) \,, 
\eena
satisfies
\begin{align}
  & 2\, k_{\text{V}}\, \Hat{D}^j {\Bbb V}_{ij}
  = \left\{ k_{\text{V}}^2 - (d - 3)\, K \right\}\, {\Bbb V}_i
  ~.
\end{align}

For massive tensor fields, it does not appear to be possible to find a single master variable for the vector-type perturbations for generic modes $k_{\rm V}$~\cite{Cardoso:2019mes}. For this reason, here we focus on the exceptional mode, 
\bena
 k_{\rm V}^2 -(d-3)K =0 \,, \quad \mbox{with $K=1$} \,.
\eena
This corresponds to the homogeneous perturbations, perturbatively adding an angular momentum. In fact, for this exceptional mode, 
$\Bbb V_{ij} $ vanishes, implying that $\Bbb V_i$ is a Killing vector field on ${\cal K}_{d-2}$. In what follows, we restrict our attention to the case of $K=1$ and the homogeneous perturbation of this exceptional mode, for the same reason as the scalar type perturbation case. 

For the exceptional mode, $H_T = 0$ identically and the $(i,j)$-components of (\ref{H_equation}) do not exist. The transverse condition becomes
\bena
 D^c \left( u^{-(d-1)/2} f_c \right) = 0 \,, 
\eena
where $D_a$ is the derivative operator compatible with the metric $g_{ab}$ of the $2$-dimensional AdS spacetime spanned by $y^a=(t,u)$ (see, the $2$-dimensional part of the metric (\ref{bdy_metric}) with $\epsilon =0$). This implies that there exists a potential $Z$ such that 
\bena
 f_a = u^{(d-1)/2} \epsilon_{ab}D^b Z\,, 
\eena
where $\epsilon_{ab} = \ell u^{-3/2} (dt)_{[a}(du)_{b]}$ is the volume element for the $2$-dimensional spacetime spanned by $y^a=(t,u)$.  

From the $(a,i)$-components of (\ref{H_equation}), we have
\bena
 D^b \left\{ u^{-d/2} D_{[a} u^{1/2}f_{b]} \right\} + \frac{\hat{m}^2}{\ell^2} u^{-(d-1)/2}f_a =0 \,.
\eena 
Substituting the expression of $f_a$ in terms of the potential $Z$ to the above equation, we have 
\bena
 \epsilon_{ab} D^b \left\{ u^{- d/2} D^c \left( u^{ d/2}D_c Z \right) - \frac{\hat{m}^2}{\ell^2} Z \right\} = 0 \,,  
\eena
which implies that the inside the curly brackets is a constant. We can set such a constant to zero by using the freedom of adding an arbitrary constant 
to $Z$: $Z \rightarrow Z + const.$, and eventually obtain 
\bena
 \partial_u^2 Z + \left( \frac{f'}{f} + \dfrac{d+1}{2u}\right) \partial_u Z +  \dfrac{1}{4u^2f^2} \left[ \hat{\omega}^2 u - \hat{m}^2 f \right] Z = 0 \,. 
\label{def:eq:vector}
\eena

In terms of $\Phi$, defined by 
\bena
 Z= u^{-d/4} \Phi \,, 
\eena
we can rewrite the above equation (\ref{def:eq:vector}) in the form of (\ref{def:eq:A}) and (\ref{def:A}) with 
\bena
 W(u) = \dfrac{1}{2\sqrt{u}f} 
          \left\{ 
                  \left(
                         p^2 -\dfrac{1}{4} 
                  \right) \frac{f}{u}   
             + \frac{ d(d+2)}{4}ff' 
          \right\} \,, 
\eena
where $p= \sqrt{\hat{m}^2 + (d-1)^2 /4}$. 
Then, choosing 
\bena
  G(u) = u^{(1+2p)/4}  \times f^{
   -(d+3+2p)/4 
  } \,, 
\eena
we find 
\bena
 {\tilde W}(u) =  \dfrac{1}{8\sqrt{u}f} \left(
   d+3+2p 
 \right)^2\,.
\eena
This is positive definite, hence $A$ is positive. Therefore, the exceptional mode in the $K=1$ case does not show instability of exponential growth.

\subsection{Tensor-type perturbations} 
The tensor-type perturbations can be given by a single scalar $\Phi$, as
\bena
  H_{ab} =0 \,, \quad H_{ai} = 0 \,, \quad H_{ij} = (u/\ell^{2})^{(d-6)/4}  \Phi {\Bbb T}_{ij}
\eena
where ${\Bbb T}_{ij}$ is the traceless-transverse tensor fields on ${\cal K}_{d-2}$, satisfying 
\bena
(\hat{\triangle} + k_{\rm T}^2){\Bbb T}_{ij}=0 \,, \quad \hat{D}^i{\Bbb T}_{ij}=0 \,, \quad {\Bbb T}^i{}_i=0 \,.
\eena
Note that the tensor type perturbations exist only for $d \geq 5$. It can be shown that $k_{\rm T}^2=0$ only for $K=0$~\cite{Kodama:2000fa}, and we assume that $k_{\rm T}^2>0$. For $K=1$, $k_{\rm T}^2$ is related to the eigenvalue $k_{\rm S}^2=l(l+d-3) , \, l=0,1,2,\dots$ of the scalar harmonics ${\Bbb S}$ as $k_{\rm T}^2=k_{\rm S}^2 -2$. So, including $K=-1$ case, we denote $k_{\rm T}^2=k_{\rm S}^2-2K=k_{\rm S}^2-2f'>0$. 
Then, from the $(i,j)$-components of (\ref{H_equation}), we obtain the equation for $\Phi$ in the form of (\ref{def:eq:A}) and (\ref{def:A}) with 
\bena
 W(u) = \dfrac{1}{2\sqrt{u}f} \left[
                                              \left\{ \dfrac{d(d-2)}{4} + \hat{m}^2 \right\} \dfrac{f}{u}
                                             + \dfrac{(d-2)(d-4)}{4} ff' + k_{\rm S}^2 f
                                       \right]\,.
\label{Vx:tensor} 
\eena

\subsubsection{Global chart $K = 1$}

Choosing $G$ as
\bena
 G = u^{(1+2p)/4} \times f^{- (d-1+2p)/4}\,, 
\eena
we find that  
\bena
 \tilde W = \frac{1}{2\sqrt{u}f} \left\{ k_{\rm S}^2 f + \dfrac{1}{4} (d-1 +2p)^2 \right\} \,. 
\eena
This is positive definite and thus the AdS in the global chart is stable against all modes of the tensor-type perturbations.    

\subsubsection{Hyperbolic chart $K = -1$}

We choose 
\bena
 G = u^{-(d-2)/4} \,, 
\eena 
we find that  
\bena
 \tilde W = \dfrac{1}{2\sqrt{u}f} 
                                       \left\{
                                                k_{\rm S}^2 f + \left(p^2-\dfrac{(d-1)^2}{4} \right) \dfrac{f}{u}
                                       \right\} 
                                           \,. 
\eena 
In the present context, $p^2-(d-1)^2/4<0$ as $\hat{m}^2<0$, and therefore $\tilde W$ can always be negative near the AdS boundary $u \rightarrow 0$ for any $k_{\rm S}^2$. 
\section{The second variation of the
${\mathcal S}_\text{bd} + S_{\text{bulk}}$
}
In the $5$-dimensional Gauss-Bonnet theory, the second variation of the boundary action 
${\mathcal S}_\text{bd} + S_{\text{bulk}}$
is given by 
\bena
\label{second_variation_bd_action}
  \delta^{(2)}{\mathcal S}_\text{bd}
+ \delta^{(2)}S_\text{bulk}
 &=& \frac{\epsilon^2}{32\pi G_5}
\int_{\p\Sigma_0} d^4x\sqrt{-\bar{\sigma}}\bar{n}_\mu V^\mu, 
\nonumber \\
 V^\mu &:=&
\frac{3}{2}H^{\alpha\beta}\BGbdyD{}^\mu H_{\alpha\beta}-H^{\alpha\beta}\BGbdyD_\beta {H^\mu}_\alpha
-\frac{24\alpha_1}{\ell^2}H^{\alpha\beta}\BGbdyD{}^\mu H_{\alpha\beta}
+\frac{18\alpha_1}{\ell^2}H^{\alpha\beta}\BGbdyD_\beta {H^\mu}_\alpha 
\nonumber \\
&{}&+2\alpha_1\BGbdyD{}^{\,2}H^{\nu\sigma}(\BGbdyD_\sigma {H^\mu}_\nu-\BGbdyD{}^\mu H_{\nu\sigma})
\nonumber \\
&{}& -4\alpha_1\left[(\BGbdyD{}^\mu\BGbdyD{}^{[\rho} H^{\alpha]\nu})\BGbdyD_\alpha H_{\rho\nu}
+(\BGbdyD{}^\nu\BGbdyD_{[\alpha} {H^\mu}_{\beta]})\BGbdyD{}^\alpha {H^\beta}_{\nu}    \right], 
\eena
where $\bar{n}_\mu$ is the unit outward normal vector defined by 
\begin{align}
\label{def_normal}
\bar{n}_\mu:=-\frac{\ell}{2u\sqrt{f}}(du)_\mu. 
\end{align}

\section{Transformation of hypergeometric function}
\bena
\label{Hyper_trans}
 F(\alpha,\beta,\gamma;z) &=&
\frac{\Gamma(\alpha+\beta-\gamma)\Gamma(\gamma)}{\Gamma(\alpha)\Gamma(\beta)}
(1-z)^{\gamma-\alpha-\beta}F(\gamma-\alpha,\gamma-\beta,\gamma-\alpha-\beta+1;1-z)
\nonumber \\
&+&\frac{\Gamma(\gamma)\Gamma(\gamma-\alpha-\beta)}{\Gamma(\gamma-\alpha)\Gamma(\gamma-\beta)}
F(\alpha,\beta,\alpha+\beta-\gamma+1;1-z), \nonumber \\
 F\left(\alpha,\beta,\gamma;z\right)&=&
\frac{\Gamma(\gamma)\Gamma(\beta-\alpha)}{\Gamma(\beta)\Gamma(\gamma-\alpha)}
(-z)^{-\alpha}F\left(\alpha,\alpha-\gamma+1,\alpha-\beta+1;\frac{1}{z}  \right)
\nonumber \\
&+& \frac{\Gamma(\gamma)\Gamma(\alpha-\beta)}{\Gamma(\alpha)\Gamma(\gamma-\beta)}
(-z)^{-\beta}F\left(\beta,\beta-\gamma+1,\beta-\alpha+1;\frac{1}{z}  \right). 
\eena


\end{document}